\documentclass[12pt,preprint]{aastex}
\shorttitle{Planetary merger}
\shortauthors{Li, Agnor, \& Lin}
\begin{document}

\title{Embryo impacts and gas giant mergers I: 
Dichotomy of Jupiter and Saturn's core mass}
\author{Shu Lin Li$^{1,2}$, C.B.~Agnor$^3$, and D.~N.~C.~Lin$^{2,4}$}
\affil{$^1$Department of Astronomy, $^4$Kavli Institute of Astronomy 
and Astrophysics, Peking University, Beijing, China}
\affil{$^2$Department of Astronomy and Astrophysics, University of
 California Santa Cruz, USA}
\affil{$^3$Astronomy Unit, School of Mathematical Sciences,
Queen Mary University of London, United Kingdom}
 
\begin{abstract} 
Interior to the gaseous envelopes of Saturn, Uranus, and Neptune,
there are high-density cores with masses larger than 10 Earth
masses. According to the conventional sequential accretion hypothesis,
such massive cores are needed for the onset of efficient accretion of
their gaseous envelopes. However, Jupiter's gaseous envelope is more
massive and core may be less massive than those of Saturn. In order to
account for this structural diversity and the super-solar metallicity
in the envelope of Jupiter and Saturn, we investigate the possibility
that they may have either merged with other gas giants or consumed
several Earth-mass proto-planetary embryos during or after the rapid
accretion of their envelope. In general, impinging sub-Earth-mass
planetesimals disintegrate in gas giants' envelopes deposit heavy
elements well outside the cores and locally suppress the convection.
Consequently, their fragments sediment to promote the growth of cores.
Through a series of numerical simulations, we show that it is possible
for colliding super-Earth-mass embryos to reach the cores of gas
giants. Direct parabolic collisions also lead to the coalescence of
gas giants and merging of their cores. In these cases, the energy
released from the impact leads to vigorous convective motion
throughout the envelope and the erosion of the cores. This dichotomy
contributes to the observed dispersion in the internal structure and
atmospheric composition between Jupiter and Saturn and other gas giant
planets and elsewhere. 
\end{abstract}


\section{Introduction} 
In the conventional sequential accretion hypothesis (SAH) for gas
giant planet formation\citep{Safronov1969, Pollack1996}, heavy
elements first coagulate into planetesimals and proto-planetary
embryos. When they attain a critical mass $M_{\rm crit} \sim 3-10
M_\oplus$, they begin to accrete gas efficiently and to evolve into
gas giants \citep{Stevenson1982, Bodenheimer1986}.

Perhaps the strongest supports for SAH are the inferred existence of
solid cores and super-solar metallicity in the envelopes of
solar-system gas and ice giants \citep{Guillot2004, Militzer2008}.
These models are constructed to match the observed masses, radii, and
moments of inertial of present-day Jupiter, Saturn, Uranus, and
Neptune.

Despite this verification, theoretical explanations are needed for two
observed properties: 1) a significant difference in the core mass
between Jupiter and Saturn and 2) the super-solar metallicity in the
envelopes of gas giants. In \S2, we discuss the uncertainties and pose
challenges to the most simple version of SAH introduced by these two
puzzles. We suggest that the dispersion in the core mass and the
enhanced metallicity in the envelope of gas giants may be the
consequence of some catastrophic collisional events which occurred well
after the onset of efficient gas accretion. We discuss the common
occurrence of giant impact and merger (GIM) events during the
formation of isolated and multiple proto gas giants as well as
throughout their long-term dynamical evolution.

We carry out numerical simulation to demonstrate that GIM's can indeed
lead to diverse planetary structures. In \S3, we briefly describe our
numerical methods and discuss the typical range of appropriate orbital
parameters and planetary properties. With a smoothed particle
hydrodynamic (SPH) scheme, we illustrate the possibility that
sufficiently massive protoplanetary embryos may be able to intrude
through the massive envelope and impact onto the cores of some gas
giants. With three examples of head-on, parabolic collisions between a
Saturn-like gas giant with super-Earth embryos or another gas giant,
we show that spherical symmetry is quickly restored after the the
cores of the merger products are thoroughly mixed.

In order to extend the simulation to a larger dynamical range in
density and over protracted phases of dynamical evolution, we
construct an spherically symmetric Lagrangian hydrodynamic (LHD)
scheme. With this LHD scheme, we consider, in \S4, impactors with a
range of masses and head-on collisional speeds. We show that 
low-velocity and low-mass impactors generally disintegrated 
in the gaseous envelope of
gas giants and their debris has a tendency to sediment towards the
planets' cores. Such events lead to the enlargement of the core
mass. However, large-mass and high-speed impactors can survive their
passage through the gaseous envelope and reach the gas giants'
core. The dissipation of kinetic energy into internal energy induces
efficient and vigorous thermal convection and erosion of the
cores. These diverse outcomes may be responsible for introducing a
variety of internal structure for long-period gas giants. In a
follow-up paper II \citep{Li2009b}, we will apply a similar scenario
to a series of models for short period planets with diverse internal
structures. Finally, we summarize our results in \S5.

\section{The diverse structure of gas giants due to giant impacts 
and mergers during proto-planetary formation}
\label{sec:section2}

In the sequential accretion hypothesis collisions represent a
principal mode of growth and planetary evolution. Giant planetary
collisions have been suggested to explain a variety of bulk planetary
characteristics in the solar system, including the origin of Earth's
Moon \citep{Hartmann_&_Davis_1975,Cameron_&_Ward_1976}, collisional
stripping of Mercury's primordial mantle
\citep{Benz_etal_1988,Benz_etal_2007} and Uranus' large obliquity 
\citep{Slattery_etal_1992}. Giant impacts during the accretion epoch display a
diversity of outcome morphologies, each with different implications
for the development of planetary characteristics
\citep[e.g.][]{Wetherill_1985, Agnor2004, Asphaug_etal_2006}.
Recent studies have begun to explore how these energetic impacts influence
the luminosity, detectability and characteristics of extrasolar planets
\citep{Anic_etal_2007, Ikoma2006}. Here we
build on these works and examine how giant planetary collisions may
account for the diversity of the interior structure of gas giant
planets, and the differences in core mass between Jupiter and Saturn
in particular. 

In this section, we provide arguments to highlight the critical roles
of giant impacts and mergers during the formation and evolution of gas
giant planets and the potential outcomes of such events.

\subsection{Observed structural diversity and its paradoxical 
implications} 
\label{sec:struct}

Although they are both classified as gas giants, Jupiter and Saturn
have very different internal structures. Within considerable uncertainty,
the estimated core mass of Saturn is more massive (by up to three times) 
than that of Jupiter. In contrast, Jupiter's total mass (which is mostly 
contained in its envelope) is around three times that of Saturn. The 
average metallicity of both planets are inferred to be several times that 
of the Sun, albeit they are unlikely to be identical.

According to SAH, gas giants acquired their cores prior to the
accretion of their gaseous envelopes. Current quantitative models
indicate that a critical core mass ($M_{\rm crit} \sim 10 M_\oplus$)
is needed for the onset of efficient accretion of the surrounding
gas \citep{Pollack1996}. This mass is determined by the heat transfer
efficiency \citep{Inaba2003} and the nebula boundary conditions
\citep{Bodenheimer2000, Wuchterl2000, Papaloizou2006, Rafikov2006}.

There are considerable uncertainties on the magnitude of $M_{\rm
crit}$. The first sets of quantitative models were constructed under
the assumptions of spherical symmetry, hydrostatic equilibrium,
efficient convection, interstellar-grain opacity, and idealized
planetesimal bombardment rate \citep{Pollack1996}. These models
indicate that uninterrupted planetesimal bombardments and a cooling
barrier can delay the formation of Jupiter in a minimum mass nebula
(MMN) by 10-20 Myr.

However, gas giants can only acquire their massive envelope inside
gas-rich protostellar disks, {\it i.e.} their formation time scale 
$\tau_m$ must be smaller than the typical gaseous disk depletion 
time scale $\tau_{\rm dep} (\sim 3-5 {\rm Myr}$). There have been 
several previous attempts to resolve this theoretical challenge. 
Alibert {\it et al.} (2004) suggested that type I migration of
Earth-mass embryos (Ward 1997) may speed up the time scale needed
for the onset of efficient gas accretion. Although this scenario
may provide a solution for Saturn (because its core is relatively
massive), it may not be appropriate for Jupiter if its core mass
is limited to a few $M_\oplus$. 

In an attempt to account for Jupiter's low-mass core, \citet{Inaba2003, 
Hubickyj2005} constructed models to show that the gas accretion rate 
increases with the heat transport efficiency. Since there are radiative
regions in proto gas giants' envelope, the critical mass for the
onset of efficient gas accretion increases with the opacity of the 
infalling envelope. The main sources of opacity in these regions 
are grains. Grain growth and formation of planetesimals can reduce
the opacity of residual disk gas substantially (see \S\ref{sec:envelo}). 
It is also possible for dust to coagulate and sediment from the 
tenuous outer envelope of proto-planets \citep{Helled2008}. 

In principle, these effects can suppress the barrier for the onset of
gas accretion. However, a rich population of super-Earths (with
masses $M_p >$ a few $M_\oplus$) has been discovered with
radial-velocity \citep{Mayor2009} and transit \citep{Queloz2009}
surveys. Some of these super-Earths have 
masses larger than Jupiter's estimated core mass (at least for some
models). This discrepancy would introduce a paradox, if we assume 
these super Earth attained their present-day mass and became ``failed
cores'' in gas-rich protostellar disks because their mass $M_p < 
M_{\rm crit}$. It is possible that the onset of gas accretion onto 
these super-Earths may be delayed by planetesimal bombardment, albeit
such events are likely to be suppressed by the formation of a gap in the 
planetesimal disk (see \S\ref{sec:embryosdyn}, \citealt{Zhou2007}).

Although this accretion time scale issue is not the central focus of
our discussion here, scenarios proposed for its resolution have
implications which exacerbate the second paradox, {\it i.e.} gas
giants' structural diversity. If the internal structure of the mature
planets is assumed to be the result of some critical conditions that
led to their formation, the large observed dispersion would imply the
existence of many possible paths for their final assemblage.

For example, both Jupiter and Saturn probably formed in a similar
solar nebula environment close to their present-day location (albeit
at different times). It is natural to assume a similar magnitude for
$M_{\rm crit}$ for the onset of efficient gas accretion, especially if
their separation may have widen through planetesimal scattering since
their formation \citep{Hahn1999}. The dichotomy of their internal
structure is indicative that it only partially reflects the critical
initial condition for the onset of efficient gas accretion. It is
entirely possible that the present-day internal structure of gas
giants has been affected by some local or temporal stochastic
processes after they have initiated (or terminated) the process of 
rapid accretion of their massive envelope.

\subsection{Heavy elemental abundance in gas giants' envelope}
\label{sec:envelo}

A related conundrum is the super-solar metallicity content of Jupiter
and Saturn's gaseous envelope. In the SAH, heavy elements are assumed
to be mostly condensed into grains, efficiently segregated from the
disk gas, and converted into planetesimals, embryos, and cores prior
to the onset of the gas accretion. The abundance distribution of
refractory elements in the meteorites suggests that the solar nebula
had an initial composition similar to that of the sun. The emergence
of solid cores with $M_p > M_{\rm crit}$ would decrease the nebula's
overall metallicity, including contribution from molecules and grains.
If gas giants' envelopes are composed solely of nebula gas and grains
accreted by their cores, they would attain a sub-solar metallicity.
This inference clearly contradicts with the observed super-solar
metallicity in Jupiter and Saturn's envelopes.

One possible solution for this paradox is that Jupiter and Saturn
formed in regions where heavy elements are locally enhanced. Due to
hydrodynamic drag, grains formed in the outer disk regions migrate
inward. They may stall and accumulate near the snow line \citep[$a_{\rm
ice}$ where the disk mid-plane temperature decreases below ice
condensation temperature at 160K, see e.g.][]{CuzziZahnle2004, Kretke2007}.
This migration barrier to grains and embryos provides a preferred location
for gas giants to assemble \citep{Ida2008b} (see \S\ref{sec:jupsat}). 
However, a strong metallicity enhancement would increase the effective 
opacity of gaseous envelopes around solid cores and the magnitude of 
$M_{\rm crit}$ \citep{Inaba2003} well beyond the current estimated core 
mass for Jupiter \citep{Guillot2004, Militzer2008}. Although coagulation
and sedimentation of grains in protoplanets' envelopes may provide a
potential solution for Jupiter's small core \citep{Helled2008}, it is
not clear why such a process did not also enable Saturn to attain a
small $M_{\rm crit}$ and low core mass.

Another possible cause of the high metallicity in gas giants'
envelopes is the erosion of their cores. However, such a scenario
would require much more massive cores (up to $\sim 30-50 M_\oplus$)
prior to the onset of efficient gas accretion. In a MMN, formation of
such massive embryos is challenging though not impossible. Core
erosion by convective gas may also be suppressed by the compositional
gradient in the boundary layer separating gas giants' cores and
envelopes \citep{Guillot2004}. This scenario can also be easily ruled
out by any dip in the planetary mass distribution in the range of
$10-100 M_\oplus$ \citep{Ida2004}.

\subsection{Emergence of cores of the first-generation gas giants}
\label{sec:jupsat}
In this paper, we propose giant impacts of residual embryos onto gaseous 
planets during and after the advanced stages of their envelope accretion
as a potential mechanism which may lead to the resolution of the structural 
diversity (see \S\ref{sec:struct}) and metal-rich envelope (see 
\S\ref{sec:envelo}). In order to distinguish between the {\it ab initial} 
core and the giant-impact scenarios, we first discuss the dynamical evolution 
of massive embryos in the context of gas giant formation.

The growth of embryos is limited by their isolation mass \citep[($M_{\rm
iso}$, see e.g.][]{Lissauer1987} and long growth timescale ($\tau_m$)
\citep{Ida2004}. In a MMN-type protostellar disks, the most massive
embryos emerge beyond $a_{\rm ice}$ within the gas depletion timescale
($\tau_{\rm dep}$) is $\sim 3-5 M_\oplus < M_{\rm crit}$. Furthermore,
the onset of gas giant formation is challenged by the type I orbital 
migration of their super-Earth embryonic cores. The rate and direction of
type I migration is determined by both the surface density ($\Sigma_g$)
and temperature $T$ (more accurately the entropy) gradient in the disk 
gas \citep{Masset2006, Paarde2010}. These quantities are 
determined by the efficiency 
of turbulent-induced angular momentum transport. Possible causes of
turbulence include the magneto-rotational (MRI) \citep{Balbus1991} 
and convective \citep{LinPap1980, LesurOgilvie2010} instabilities.

Near the snow line of a MMN, the ionization fraction near the disk midplane
is negligible and the gas there is unaffected by MRI turbulence. However,
the ionization fraction may be much higher near the disk surface layers
which are exposed to the ionizing photons from the central stars and 
elsewhere \citep{Gammie1996}. The depth of layers which are affected by 
the MRI turbulence is determined by an ionization equilibrium in which 
the ionization rate is balanced by the rate of recombination on the 
grains. Interior to $a_{\rm ice}$, the sublimation of volatile ices 
eliminates a majority of the recombination sites, enhances the 
ionization fraction of the disk gas, and increases the depth of the 
surface layer which is regulated by MRI turbulence. Across $a_{\rm ice}$,
effective viscosity has a negative gradient, which in quasi steady
state, leads to a positive gradient in $\Sigma_g$ and $P_g$. For a
range of disk mass accretion rates ($\dot M_d \sim 10^{-9} - 10^{-8}
M_\odot$ yr$^{-1}$), the local $P_g$ maximum at $a_{\rm ice}$ provides
a natural barrier for the orbital decay of grains against hydrodynamic
drag \citep{Kretke2007}. This structure also reverses the direction of
type I migration due to the effect of horseshoe drag \citep{Masset2006}. 
Outward migration of embryos by a few AU may also be induced by an 
entropy gradient \citep{Paarde2010}.

Thus, similar to grains' orbital evolution, embryos' migration leads 
to them to congregate near $a_{\rm ice}$ where both the gas surface 
density ($\Sigma_g$) and pressure ($P_g$) attain local maxima in the disk 
due to the transition in the efficient turbulent angular momentum 
transport. Through the accumulation of neighboring embryos and
planetesimals a protoplanet may acquire sufficient 
mass ({\it i.e.} $\sim M_{\rm crit}$) to evolve into a core which
efficiently accretes gas \citep{Ida2008b}.

\subsection{Embryos' dynamical evolution around isolated gas giants}
\label{sec:embryosdyn}

During the early phase of gas accretion, the time scale for a
proto-planet to double its mass is longer than the synodic periods
(relative to its orbit) of planetesimals and embryos within its
feeding zone. Its tidal perturbation on the nearby embryos and their
interaction with the disk gas induces their orbit to evolve away from
that of the dominant proto-planet \citep{Zhou2007, ShiraishiIda2008}. 
Therefore slowly (``adiabatically'') growing proto-planets cannot
accrete those embryos within their expanding feeding zones in contrast
to the assumption made in previous models \citep{Pollack1996,
DodsonRobinson2008}.

This initial suppression of giant impacts reduces the energy released
by the growing proto-planet and promotes the gas accretion rate. In
the limit of negligibly small thermal feedback and weak
proto-planetary torque, proto-planets' uninhibited gas accretion rate
($\dot M_p$) may be approximated by the Bondi formula such that the mass
doubling time scale would be
\begin{equation}
\tau_m \equiv{M_p \over \dot M_p} \sim \left({M_\ast \over \pi
\Sigma_g a^2} \right) \left( { M_\ast \over M_p} \right) \left( {H
\over a} \right)^4 {P \over 2 \pi}
\end{equation} 
where $H$ is the disk thickness. 

The above expression indicates that, once it is initiated, Bondi-type 
gas accretion is a runaway process. During their growth, there are several
effects which can lead to GIM events:

\noindent
A1) Proto gas giants' accretion rate $\dot M_p$ is eventually
suppressed by the structural changes in their natal disks induced by
their tidal perturbation when their $M_p$ exceeds $\sim (H/a)^3
M_\ast$, which is comparable to $M_J$ in a MMN near $a_{\rm ice}$
\citep{Dobbs-Dixon2007}. However, just before they have acquired an
adequate mass to induce gap formation, proto-planets' $\tau_m$ reduces
below a thousand years \citep{Tanigawa2002}, which is comparable to
the synodic periods of their nearby planetesimals and embryos (with
semi major axis $a \pm \delta a \sim 0.1-0.2 a$). Gas giants' rapid
(non adiabatic) mass increases destabilize these embryos' orbits and a
modest fraction of these objects collide with and are captured by the
emerging gas giants \citep{Zhou2007}.

\noindent 
A2) After gas giants' accretion is quenched by gap formation, their
gravity perturbs the orbital elements of residual planetesimals and
embryos out to quite large distances. To the lowest order, gas giants'
secular perturbation leads to the modulation of the embryos' eccentricity
and the precession of their longitude of periastron. The amplitude of
embryos' eccentricity modulation increases with gas giants'
eccentricity $e_p$ and $M_p$ \citep{Murray1999}. Many known gas
giants have significant $e_p$'s and they can induce modest amplitude
eccentricity modulations among their nearby embryos.

Embryos' orbital precession frequency is determined by their
$a$'s. Due to the self gravity of their natal disks, orbits of gas
giants and residual embryos also precess with frequencies determined
by their $a$'s relative to the $\Sigma_g + \Sigma_d$ distribution of
the disk. Secular resonances occur at special locations where the
embryos' precession frequencies due to the gravitational perturbation
of the gas giants matches their differential precession frequencies
due to the disk potential. In these resonances, embryos'
eccentricities can be greatly excited by the nearly constant
differential longitude of periastrons. During the depletion of the
disk, gas giants' secular resonances sweep over wide regions and
destabilize the orbits of the planetesimals and embryos at these
locations \citep{Ward1981, Nagasawaetal2005, Thommes2008}. Some
embryos may cross the orbits of and collide with the gas giants.

\subsection{GIMs in systems with multiple gas giants.} 
\label{sec:multiplanet}

There are two gas giants, Jupiter and Saturn, in the solar
system. Several stars (such as $\upsilon$ And) have known multiple
planets around them. Around other planet-bearing stars, there are
indications of additional planetary companions with comparable $M_p$'s
\citep{Cumming2008}. In these systems with multiple planets, there are
several additional avenues which may lead to catastrophic impacts of
residual embryos onto their envelopes. For example:

\noindent A3) Regardless of the low formation and survival probability
of the first-generation planets \citep{Ida2008b}, their birth leads to
gap formation which provide a secondary pressure maximum in the gas
disk outside their
orbits \citep{Lin1979, Lin1993, Bryden2000a}. This induced disk
structure also sets up a barrier against the orbital
decay of grains and embryos. The accumulation of heavy elements beyond the outer edge of the
gap promotes the formation of more-distant second-generation cores on
time scales much shorter than that required for them to assemble in
isolation \citep{Bryden2000b}. In this sequential formation scenario,
the second-generation gas giants' rapid mass increase, during their
advanced growth phase, also introduces non-adiabatic changes in the
systems' gravitational potential which have a tendency to destabilize
the orbits of the nearby residual embryos.

\noindent A4) After the gas depletion, planets' orbits continue to
evolve due to their scattering of residual planetesimals. If multiple
planets' orbits have spread out from a more compact configuration,
orbital instabilities and scattering amongst giant planets may also lead to
late accretion of planetesimals by gas giants \citep{Tsiganis2005}.

\subsection{Disruption of impacting embryos} 
\label{sec:disruption} 
The main focus of this paper is to examine the structure adjustment of
gas giants' envelopes after major merger events rather than a 
statistical study on the coalescence probability. We do not 
consider collisions in which the intruding embryos pass through 
the envelope of gas giants without losing a significant fraction 
of their heavy elemental masses. For merger models, we choose 
representative rather than a realistic 
distribution of models to illustrate the possibility and character 
(rather than probability) of two sets of 
outcomes, {\it i.e.} the intruding embryos either disintegrate in 
the gas giants' envelope or impact onto their cores. 

For simplicity, we consider mostly head-on collisions. Impacting planetesimals
much smaller than the giant planet would likely suffer ablative
disintegration during their passage through the gaseous envelope.
On the other hand sufficiently massive impactors may penetrate deep
and reach the core. A projectile's disintegration becomes likely when
it collides with an air-mass comparable to its own
\citep{Korycansky2005}. For head-on 
collisions, the distance traveled in a gas giant's envelope before
disintegration is approximately $D \sim (\rho_e/\rho_g) R_e$ where $\rho_e$ and
$\rho_g$ are the average density of the embryo and gas giant, and
$R_e$ is the radius of the embryo. Reaching the center of a planet
($D\gtrsim R_p$) requires a sufficiently large or massive impactor
($R_e/R_p\gtrsim (\rho_g/\rho_e)^{3}$). For Saturn-like gas giant
planets ($\rho_g=0.5-1.5$ g cm$^{-3}$) hit by condensed rocky embryos
($\rho_e=4-6$ g cm$^{-3}$) reaching the core becomes feasible for
impactors of $\sim M_{\oplus}$ or larger. We note that the
length of the path to the core, the air-mass encountered and hence the
embryo mass required to reach it all depend on the impact orientation. Further, the depth at which
an impactor's mass and energy is deposited is sensitive to the impact
orientation \citep{Anic_etal_2007}.

\subsection{Mergers of gas giant planets}
\label{sec:gasgiantmerger}
In relatively massive disks, the formation of multiple systems may
lead to merger events due to the following effects:

\noindent
B1) The emergence of the second-generation gas giants can perturb the
orbit of their predecessors and induce them to undergo type III
runaway migration \citep{Masset2008, Zhang2007}. During this phase of
rapid orbital evolution, planets' orbits may be dynamically
destabilized even though they are surrounded by considerable amount of
disk gas (\citealt{Zhou2007}, also see \S\ref{sec:embryosdyn}).

\noindent
B2) After the depletion of the disk gas, long-term dynamical
instabilities induced by all the gas and ice giants on each other (and
the embryos) can lead to their eccentricity excitation and orbit
crossing \citep{Duncan1987, Zhou2007b} on a time scale
\begin{equation}
{\rm log} (T_c/P_k) = -5 + 2.2 k_0
\label{eq:tcross}
\end{equation}
where $P_k$ is the mean orbital period of the planets, $k_0 \simeq 2 (q
-1) a_i/ (q+1) r_{\rm H}^\prime$, $q = a_{i+1} / a_i$ is the ratio of the
semi major axis between planets $i$ and $i+1$, and the modified Hill's
radius $r_{\rm\ H}^\prime = (M_p (a_i) + M_p (a_{i+1}) /3
M_\ast)^{1/3} a_i$ takes into account the mass of two interacting
proto-planets\citep{Zhou2007b}.

\noindent
B3) Eccentricity growth and orbit crossing widen planets' feeding zones
but reduce their collisional cross sections to its geometric values.
Close encounters can eventually lead to velocity dispersions
comparable to the planets' surface escape speed $V_{\rm esc}$.
Interior to the region where the planets' Keplerian orbital velocity
$V_{\rm kep} \sim V_{\rm esc}$, they undergo repeated close encounters
until they eventually collide with each other \citep{Lin1997}.

\section{Models of giant impacts onto gas giant planets} 
\label{sec:3} 

Having established that GIMs are expected to
commonly occur among long-period gas giants, we use two methods to simulate
giant impacts between gas giants and planetesimals and mergers between two
planets. In the first method, we model collisions into gas giant planets using
smoothed particle hydrodynamics (SPH). It is a Lagrangian particle
method suitable for modeling highly deformed flows and shocks evolving
in three dimensions. Because it only tracks material particles of
interest, modeling flows into empty space can be done
economically without the need for including vast volumes of
all potentially occupied space
as in a grid type code. This versatile boundary-free multi-dimensional 
hydrodynamic scheme is well suited for the simulations of colliding
astrophysical and planetary bodies. 

A few dynamical times after a giant impact, the gross
deformation gravitationally settles back to a new roughly symmetrical
equilibrium shape. Over short timescales and for densities near that
of the planet, the SPH model provides a good description of the
planetary response. We examine the long term thermal adjustment of the
gaseous envelope using a 1D (spherically symmetric) Lagrangian
radiative hydrodynamic scheme. 
 

\subsection{3-D Smoothed particle hydrodynamic models}
\label{sec:sphmethod}

SPH is widely used for modeling flows in a variety of
astrophysical and planetary contexts \citep[see e.g.][for reviews of
the method]{Monaghan1992, Benz1990}.
It has been used for a variety of problems including modeling stellar
collisions \citep[e.g.][]{Benz1987} as well as impacts between condensed solar system
objects \citep{Benz1986, Love1996, Benz1999, Asphaug1998}. The model we use
here employs a hierarchical tree to compute self-gravity and is a
variant of the code used in \citet{Benz1986}.  


To simulate impacts into gas giant planets with cores, we must model the
hydrodynamic response of both condensed and gaseous matter. We construct model
planets using the Tillotson \citep{Tillotson1962} equation of state (see
Appendix II of \citealt{Melosh1989}) with the material parameters of iron to
simulate the structure of the condensed core. The algebraic expression of this
equation of state is convenient for use in SPH schemes.

We treat the giant planet's gaseous envelope using a perfect gas equation of
state, i.e. $P_g = (\gamma-1)\rho u$ where $P_g$, $\rho$ and $u$ are the
pressure, density and specific internal energy respectively and we have used
$\gamma=5/3$. Gas giant planetary models are constructed by surrounding an
iron core with a gaseous envelope and letting the system relax to a state of
hydrostatic equilibrium. In regions with modest density, the smoothing length
is adjusted to maintain a specific number of neighbors. For numerical
convenience, we employ a lower limit on the density of $\rho_l = 0.5 g
cm^{-3}$ such that the expansion of the smoothing length is suppressed for
particles with density below this value. This density ``floor'' prevents gas
in the individual fragments (or the envelope) from dispersing and diffusing
out to tenuous medium. Since there is limited envelope expansion and very 
little mass loss associated with models presented here, this numerical 
prescription does not affect our results. 


Finally, we model large planetary impactors 
using the Tillotson material parameters for both iron and basalt. In 
principle, impactors might be primarily composed of water ice. 
Impactors' composition for a given mass can determine whether they 
disintegrate or impact onto the core.  As we are simply interested
in illustrating the gross characteristics of merger events (rather than 
numerically measuring the critical impactor mass or impact parameter
which determine specific outcomes) our main conclusions are
not strongly tied to the particular material properties chosen
(e.g.~basalt, iron, ices) or the use of head-on vs the more common
oblique collisions for particular impact models.  




As discussed in \S\ref{sec:section2}, impactors into gas giant planets
range from planetesimals and embryos to full grown planets. To
explore a few of these possibilities, we have performed three SPH simulations of
impacts into a `Saturn-like' gas giant planet of
mass $M_T = 100M_{\oplus}$ which includes an iron core with a mass $M_{T,c} =
10 M_\oplus$. The first and second models are designed to consider the
effects of giant impacts onto a gas giant planet by a planetary impactor with
$M_I = 10 M_\oplus$ and $M_I = 25 M_\oplus$ respectively. The impactor is
assumed to be entirely made of basalt such that its core mass $M_{I,c} = M_I$.
A third model examines the merger of two gas giants each with $M_T =
100M_{\oplus}$ and $M_{T,c} = 10 M_\oplus$.
\label{sec:sph}


For long-period planets, it's not likely for the high-velocity collisions to
occur. Therefore, we set the impact speed $v_{\rm imp} = v_{\rm esc}$, where
the two-body escape velocity is $v_{\rm esc} = (2G (M_I + M_T)/(R_I +
R_T))^{1/2}$ for the models in this paper. 
Model parameters and results of these simulations are
summarized as ``S'' series in Table \ref{tab:largea}.

\begin{table}
\begin{center}
\begin{tabular}{cccccccccc}
\hline
 Model & $M_{T}$ & $M_{T,c}$ & $M_{I}$ & $M_{I,c}$ & $M_{f,c}$ 
 & $M_{f,g}$ & $R_f/R_i$ & \\
 & $(M_{\oplus})$ & $(M_{\oplus})$ & $(M_{\oplus})$ & $(M_{\oplus})$
 & $(M_{\oplus})$ & $(M_{\oplus})$ & & \\
\hline \\
 Sa & 100 & 10 & 25 & 25 & 35 & 90 & 1.08 \\
 Sb & 100 & 10 & 10 & 10 & 20 & 90 & 1.06 \\
 Sc & 100 & 10 & 100 & 10 & 20 & 180 & 1.25 & \\
 La & 100 & 10 & 25 & 25 & 35 & 90 & 1.02 & \\
 Lb & 100 & 10 & 10 & 10 & 20 & 90 & 0.996 & \\
 Lc & 100 & 10 & 100 & 10 & 20 & 180 & 1.59 & \\
\hline 
\end{tabular}

\caption{Simulations of collisions between the Saturn-like gas giant and
embryos with masses of 25 $M_{\oplus}$ and 10 $M_\oplus$, and another
identical Saturn-mass gas giant, which are denoted as model a, b and c
respectively. The models named as ``S'' correspond to the SPH calculations,
while those named as ``L'' correspond to the one dimensional LHD calculations.
All the models are for parabolic, head-on, low-speed collisions. For the
composition of the impactors, the embryos are made of Tillotson basalt, and
the gas giant contains a 10-$M_\oplus$ iron core and a 90-$M_\oplus$ gaseous
envelope. Quantities $M_{T}$, $M_{T,c}$, $M_{I}$, $M_{I,c}$, $M_{f,c}$, and
$M_{f,g}$ correspond to the target' total mass, its core mass, impactor's
total mass, its core mass, the planet's final mass in the core and gaseous
envelope respectively. Planets' initial and asymptotic radii when 
a quasi hydrostatic equilibrium is re-established after the GIMs
are represented by $R_i$, and $R_f$ respectively. 
\label{tab:largea}}

\end{center}
\end{table}

\begin{figure}
\includegraphics[width=0.5\textwidth]{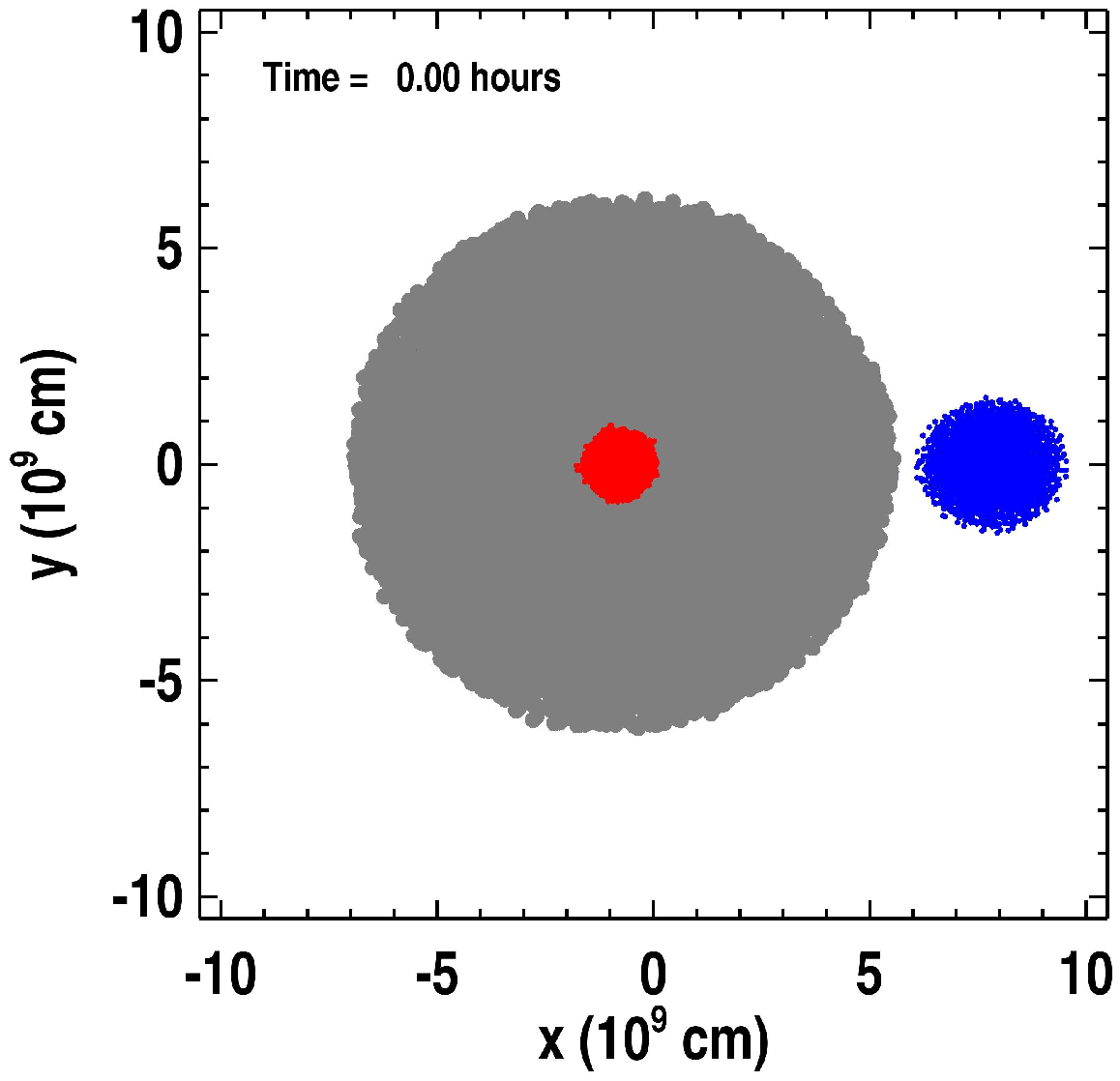}
\includegraphics[width=0.5\textwidth]{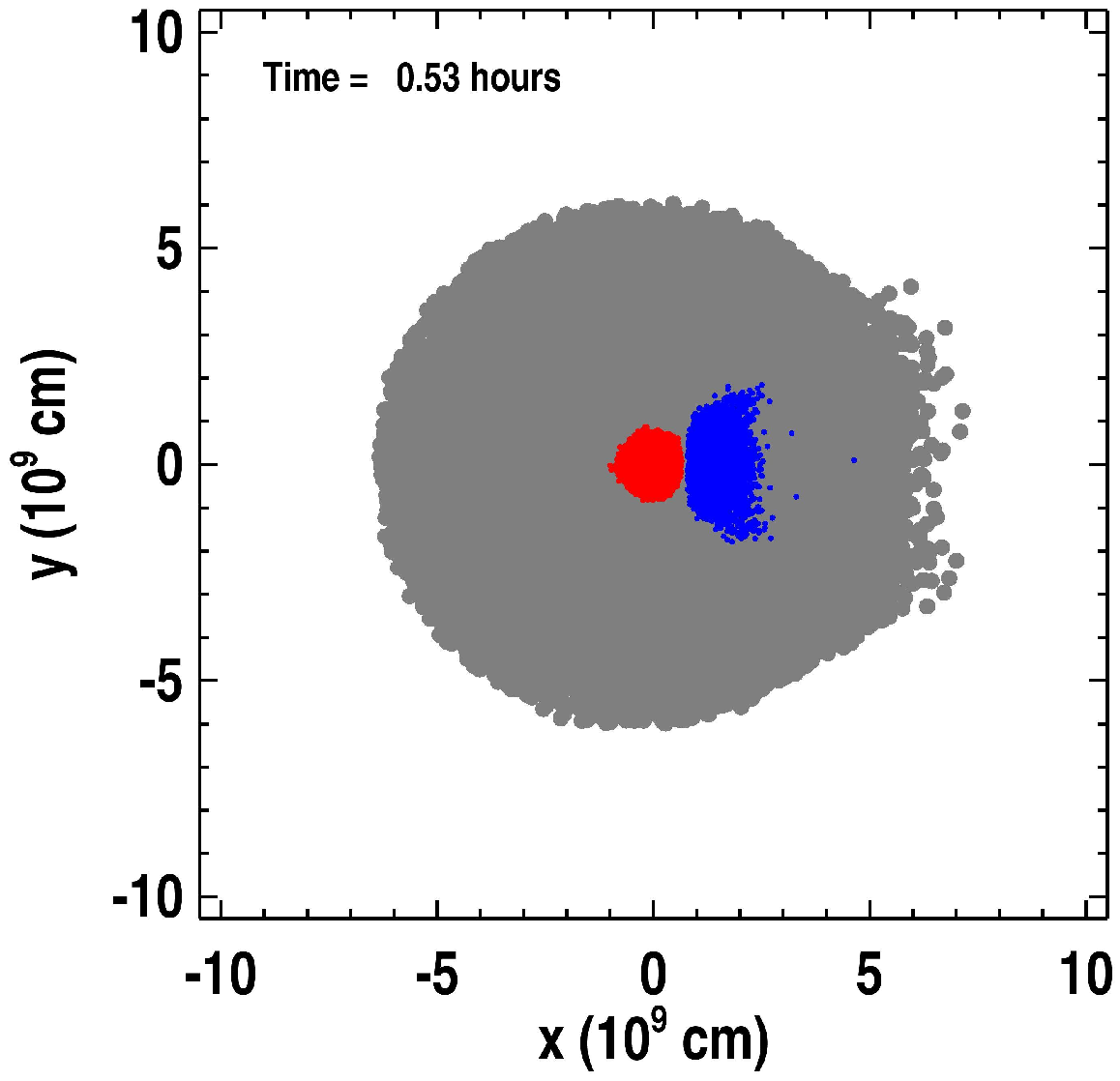}
\includegraphics[width=0.5\textwidth]{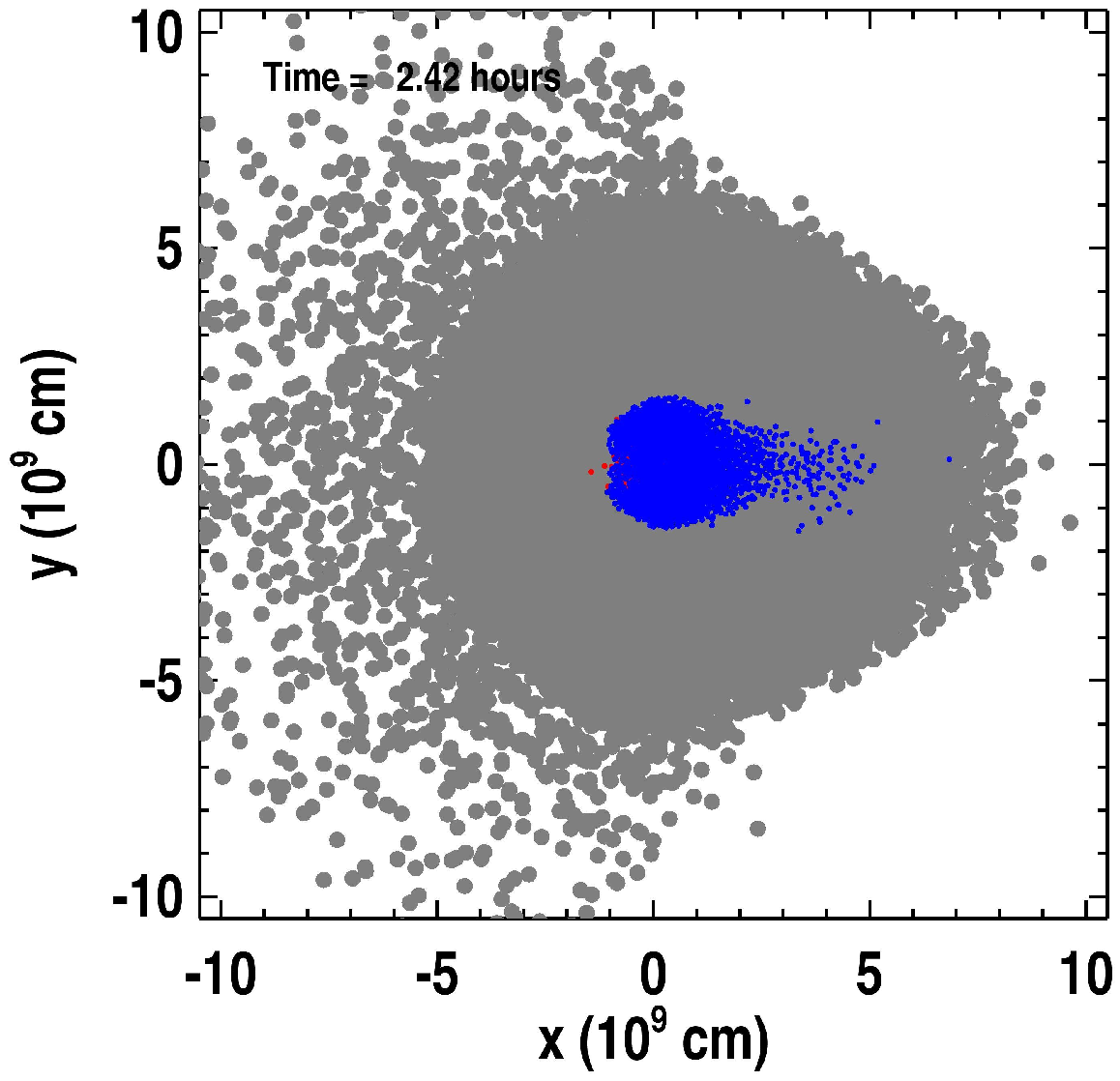}
\includegraphics[width=0.5\textwidth]{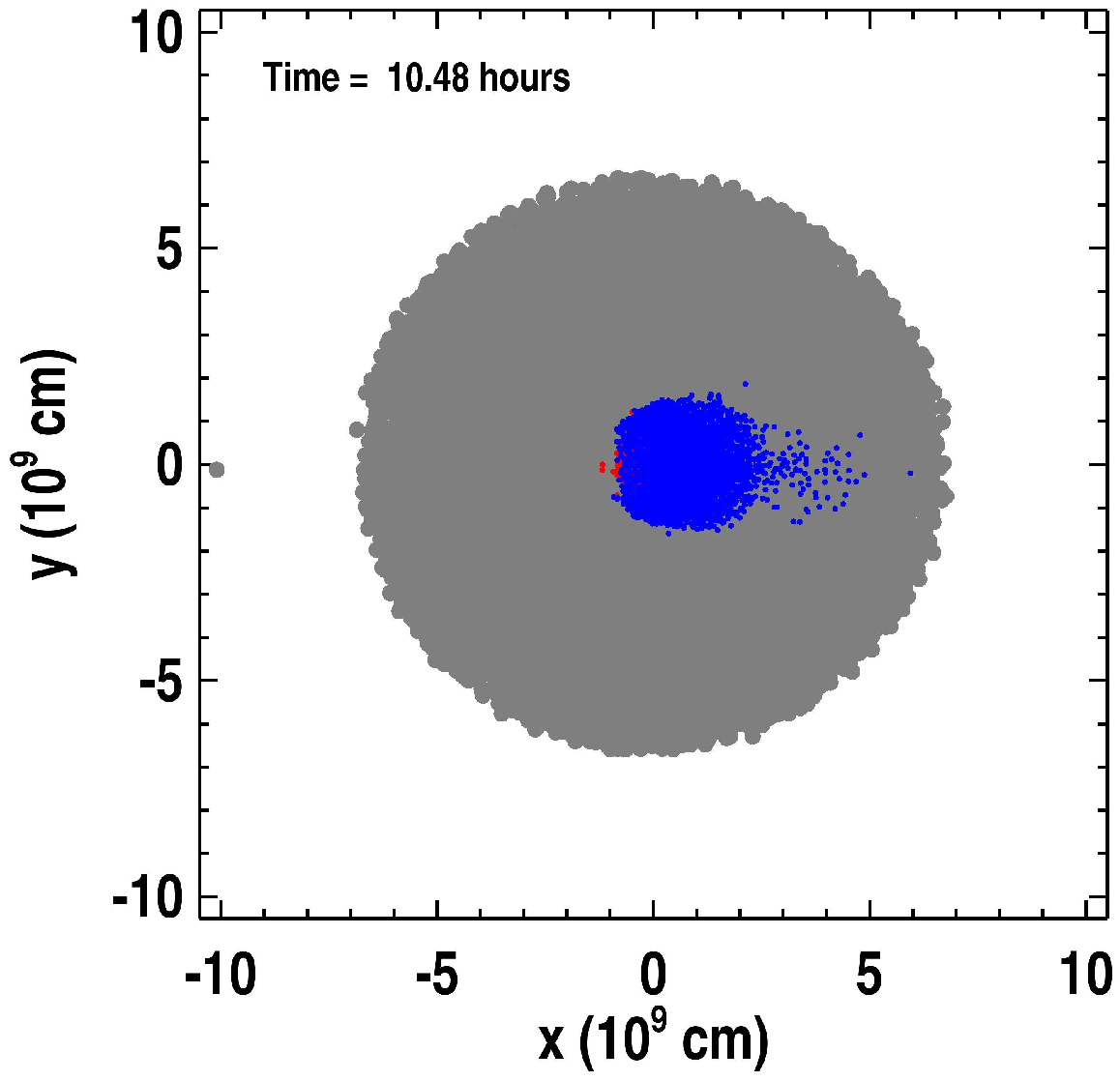}
\includegraphics[width=0.5\textwidth]{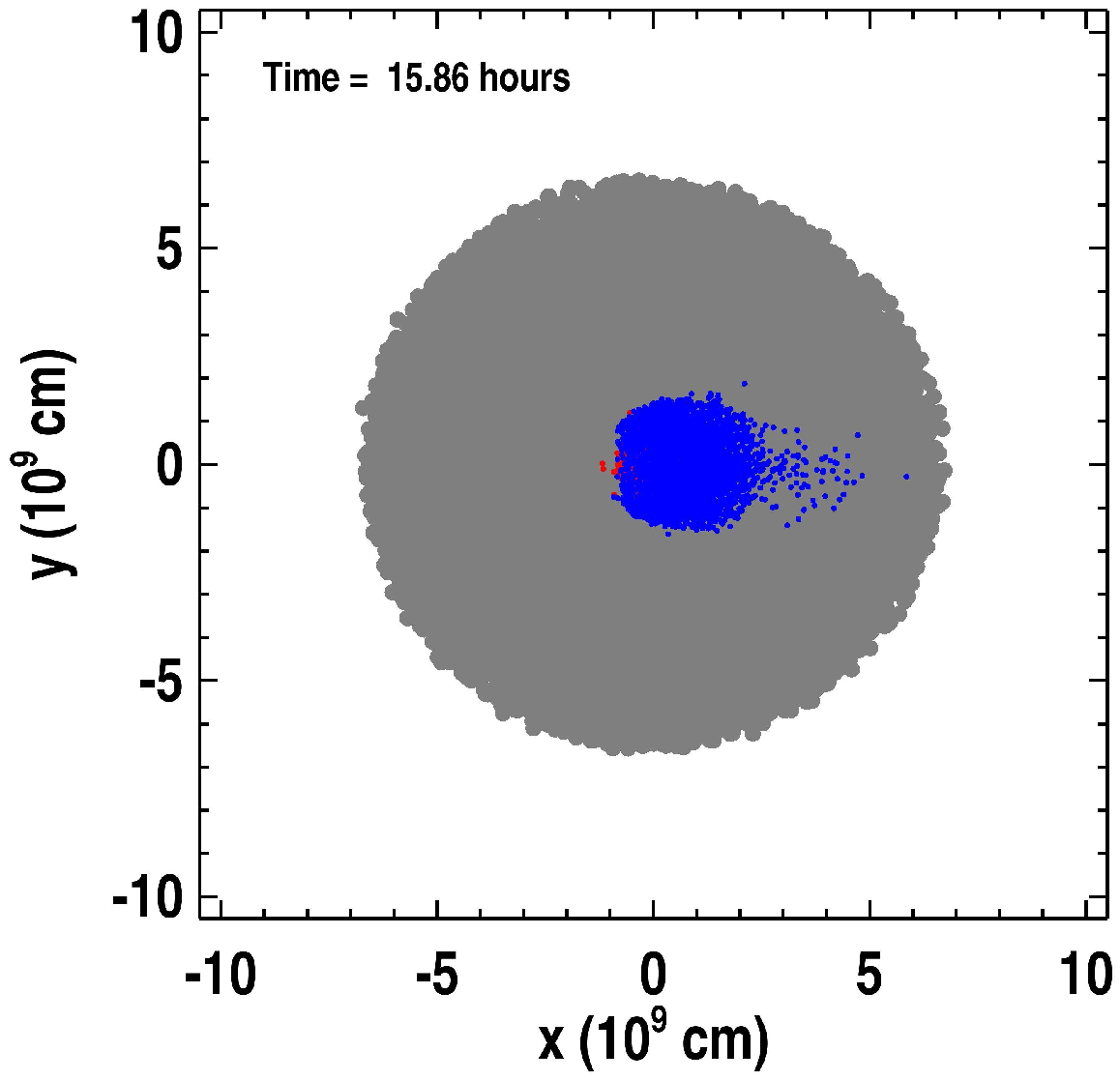}
\includegraphics[width=0.5\textwidth]{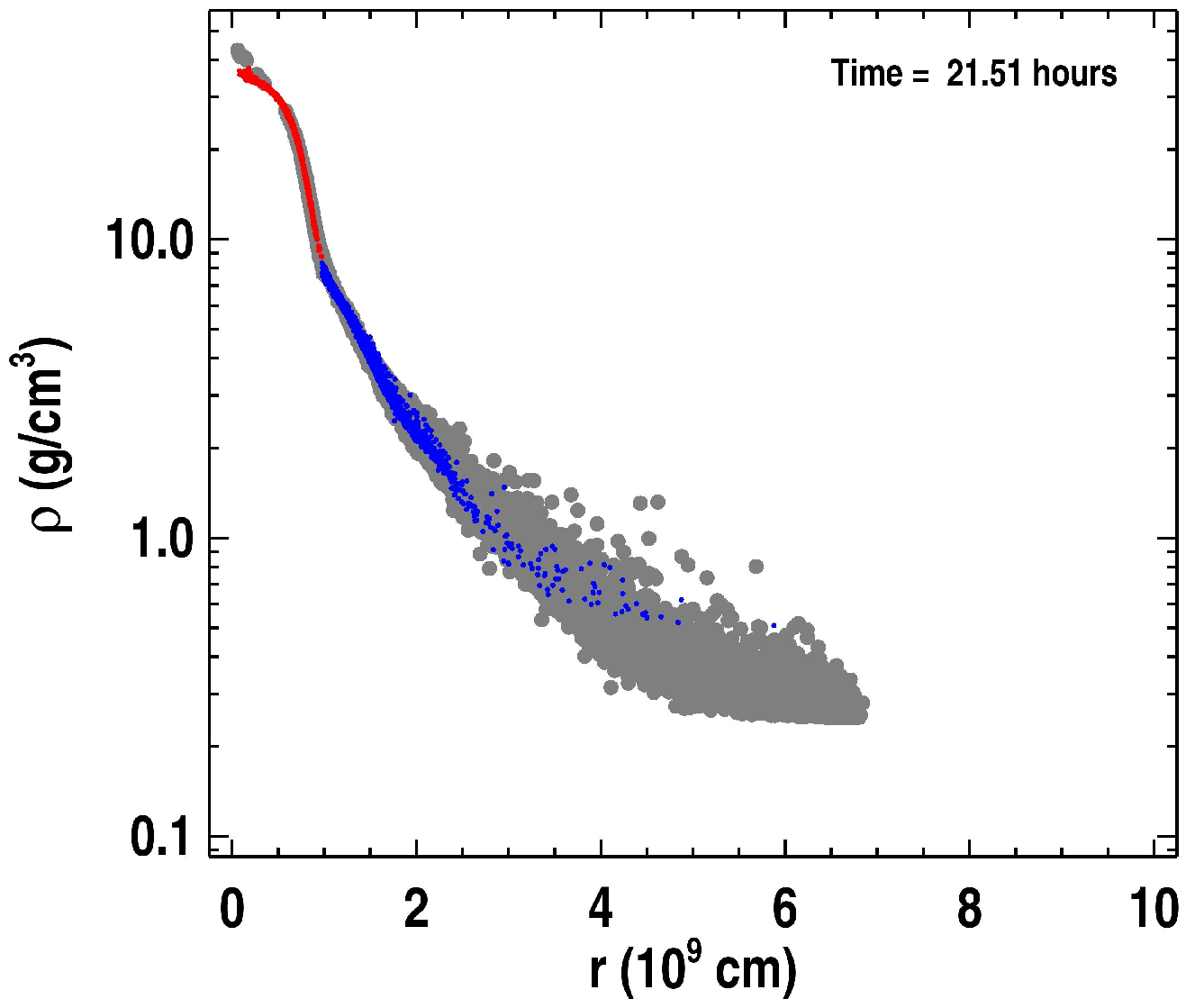}
\caption{Collision between a Saturn-like gas giant and a 10-$M_\oplus$
 impactor (Run Sb). Gas particles are shown as grey. Core particles are
overplotted in red, and the basalt impactor is overplotted in blue.
\label{fig:modelSb}} 
\end{figure}

As expected in relatively low-speed, parabolic collisions, the impactor always
merges with and is consumed by the gas giants. The gross morphology of the
collision product and the delivery of the impactor mass and energy into the
giant planet are illustrated in snapshots of the Sb model shown in Figure
\ref{fig:modelSb}.

During the initial stages of collision the impactor excavates a
crater in the gas giant's envelope. Over a few dynamical times the
initial deformation propagates throughout the planet the planet and body
begins oscillating globally and relaxing as the impactor travels to
planet's center. In the course of this passage, the impactor is
flattened and sheds mass before eventually reaching the core. As
discussed above, a 10-$M_{\oplus}$ impactor is sufficiently massive
that a significant fraction of its initial mass reaches the core at once.
The impactor then equilibrates with the core, material settles back
into a new state of hydrostatic equilibrium and the planet
assumes the expected larger spherically symmetric shape. 
 
A portion of the impactor and impact energy are deposited high up in the
gaseous envelope. Eventually sedimentation of this mass may release
additional energy on a much longer time scale. While the precise
partitioning depends on impactor mass, velocity and orientation, it is
clear that large impacts can deposit mass and energy in large parcels
into both the core and envelope.

In addition, the final volumetric radius of the SPH result is in good agreement
with those of our one-dimensional LHD model (see below) suggesting
that longer term thermal evolution may be encapsulated using a
more computationally convenient approach.

\subsection{The 1D Lagrangian radiative hydrodynamic (LHD) impact model}
\label{sec:lhdmethod}

The 3D-SPH scheme is not ideally suited for the demanding computation of the
long-term evolution of the gas giant planets. 
The results generated from the preliminary SPH
scheme indicate that internal flow quickly isotropizes the density distribution
of the merger product (see below). In the limit of negligible rotation, the
structural response and evolution of the envelope, due to the release of
thermal energy during the core coalescence, can be computed with a 1D
Lagrangian hydrodynamic (LHD) scheme. This approach was firstly adopted by
Wuchterl \citep{Wuchterl1991} in his calculation of the planet formation
problem. Although a similar approach is used, our independently developed LHD
scheme is based on a more conventional prescription for the convective heat
transfer and an alternative approximation for the computation of the thermal
evolution of planets (see \S\ref{sec:convec}).

\subsubsection{Governing equations for the LHD scheme}
\label{sec:LHDeqn}

The basic equations for the LHD scheme include three conservation laws
of momentum, mass and energy, and the equation of state for the
relationship among the thermodynamical quantities. When writing them
in the Lagrangian scheme, these equations are as follows:

\begin{equation}
{\partial u\over {\partial t}}=-{1\over \rho_0}\left({R(r,t)\over r}\right)^2
{\partial (P+q)\over \partial r} + {\partial \Psi \over \partial r} 
\label{eqn:1}
\end{equation}

\begin{equation}
{\partial R\over \partial t}=u
\label{eqn:2}
\end{equation}

\begin{equation}
{1\over \rho} ={1\over \rho_0}\left({R(r,t)\over r}\right)^2
{\partial R\over \partial r}
\label{eqn:3}
\end{equation}
 
\begin{equation}
{\partial \varepsilon \over {\partial t}}=\left({{P+q}\over {\rho ^2}}\right)
{\partial \rho \over {\partial t}}-{1\over \rho} \nabla \cdot \overrightarrow{F}
\label{eqn:4}
\end{equation}

\begin{equation}
\varepsilon =f(P,\rho)
\label{eqn:5}
\end{equation}
where the symbols denotes their usual meanings. Note that $r$ denotes the
element coordinate at some particular time which corresponds to the fluid
density $\rho_0$. Usually, these two values are viewed as our initial
condition. While $R(r,t)$ means the coordinate at time $t$ for the same
element which initially located at $r$. In order to smooth the shock, we
included the artificial viscosity denoted as $q$. Dissipation associated with
the artificial viscosity is added to equation (\ref{eqn:4}) to ensure the
conservation of total (kinetic, potential, and internal) energy.  Since
we are primarily interested in cohesive collisions and mergers with no mass
loss, the gravitational potential includes only planet's contribution.



When discretizing the equations, we use a staggered grid in space. As the code
is written in Lagrangian scheme, the coordinates will change with the gas
flow. Thus, the interval of the grids may become very small (for example, in
the piling up region) or become very large (when the gas is heated by the
giant impact and expands). That would not only reduce the time step largely,
but also make the differentiating between the grid inaccurate. The solution is
to readjust the mesh after every time step, i.e. merging the nearby two grids
when the interval becomes small, and splitting one grid into two when the
interval becomes unacceptably large. 

During the initial evolution after each major impact, we solve the continuity
and momentum equations explicitly. The time steps for the numerical
integration is chosen appropriately so that the Courant condition is
satisfied. In order to avoid numerical instability, we implicitly solve for
the energy equation, which can be converted to the form:

\begin{equation}
-c_v{{\partial T}\over {\partial t}}+
{{P\delta}\over {\rho ^2\alpha}}{{\partial \rho}\over {\partial t}}=
{1\over {\rho }} \nabla \cdot \overrightarrow{F} 
\label{eqn:energy}
\end{equation}

\subsubsection{Thermal energy transport}
\label{sec:convec}

In the SPH scheme, the equation of state for the gas is approximated by a
$\gamma=5/3$ polytrope and also by an ideal gas equation of state. 
Kinetic energy is dissipated into thermal energy during the impact.
Subsequently, the thermal energy is transported to the surface and the merger
product contracts on a Kelvin-Helmholtz time scale.

One advantage of the LHD scheme is its versatility which allows us to
incorporate various prescriptions for the equation of state and consider the
influence of phase transition, elemental separation, density and molecular
weight distribution. These effects determine whether convection may be
stabilized in some regions of the envelope. The \citet{Saumon1995} equation
of state is used, which especially takes into account the non-ideal effects of
hydrogen at high pressure.

For the transported energy flux $F$ in Eq.~(\ref{eqn:4}), we considered both
radiative and convective energy transfer. In the radiative regions, $F$ is
calculated using the radiative diffusion equation with a flux limiter
\citep{Burkert2005}. For opacity, we include the contribution from grains
which is generally several orders of magnitude larger than that from gas
molecules \citep{Alexander1994}.

In our LHD, convection is calculated using the standard mixing length
prescription \citep{Kippenhahn1990}. In the quasi hydrostatic planetary
structure models developed by \citet{Pollack1996}, convection is assumed to be
efficient and the gas attains an adiabatic state throughout the envelope. This
assumption is adequate in the envelope of present-day gas giants where the
density is sufficiently high for convection to be effective. However, in the
simulation of giant impacts, the envelope may expand greatly to attain a low
density state in which convection is unlikely to be efficient (see Figure
\ref{fig:thermal}). Therefore, we adopt the more
realistic mixing-length treatment for the convection during both expansion
and quasi steady contraction phases. This effect is important in determining
the rate of radiation transfer, the extent of envelope expansion, and the rate
of planetary contraction.

In contrast to the prescription developed by \citet{Wuchterl1991}, we apply
the standard mixing length prescription under the assumption that the
background gas is in a quasi-hydrostatic equilibrium state. This
approximation is adequate for the stage during which the magnitude of the
radial velocity of the gas ($\vert u \vert$) is smaller than the convective
speed $V_{con}$. However, during the dynamical transition phase after a giant
impact, the expansion of the envelope is supersonic over some regions of the
envelope and the efficiency of convection may fall below that estimated with
the mixing length prescription. Although the validity of the mixing length
prescription is weakened, it does provide an upper limit on the rate of energy
loss and a lower limit on the extent of the envelope's expansion.





\subsubsection{Initial parameters for the LHD models}

We assume the planet is initially in a hydrostatic equilibrium. In order to
construct an initial model for a gas giant planet, we use Runge-Kutta
integration method to generate the internal structure of the planet. In these
calculations, constant entropy is assumed to represent efficient heat
transport in the convection zone.

The initial model is relevant for a planet with a total mass $M_p = 100~
M_\oplus$ and radius $R_p = 6.2\times 10^{9}~cm$. It contains a core with
$M_{\rm core} = 10~M_\oplus$ and $R_{\rm core} = 2.2~R_\oplus$. The surface
temperature of this planet is set to be $T_e = 150$ K. Such a model
corresponds to a long-period gas giant planet with an equilibrium temperature
at around 5 AU from a solar-type star. These initial conditions are
similar to those used in the SPH simulations.

Starting with these initial conditions, we further evolve the model for a few
years and compare our results with those obtained by planetary structure and 
evolution method (under the assumption of hydrostatic equilibrium, cf. 
\citet{Pollack1996}). This computed
duration is long compared to the dynamical time scale but short compared with
Kelvin Helmholtz time scale. Although the results obtained with these two
prescription agree well in the dense inner envelope, they differ slightly in
the outer tenuous regions. The main cause for this difference is that
convection is unlikely to be efficient in the low density outer regions.
Although such an approach is more time consuming, we adopt a more realistic
mixing length prescription for our LHD models.

\section{LHD simulations of Giant Impact and Merger of Gas Giants}
\label{sec:lhd}

The results of SPH simulations show complex flow pattern during giant impacts
of embryos with gas giants and during mergers of two gas giant planets.
In a few dynamical times, the structure of the core and gaseous
envelope quickly restores to spherical symmetry during head-on
collisions. In this limit, we utilize the LHD scheme to study the
evolution of the GIM products' structure over a larger dynamical range. 



In this series of simulations, we adopt the same model parameters as we used
to simulate head-on parabolic collisions with the SPH scheme. As
expected (and illustrated in \S\ref{sec:sph}), after  initial
transitions, spherical symmetry is rapidly re-established.
SPH simulations also
indicate that relatively large proto-planetary cores can survive
passage through the gas giant's envelope and merge with the core. Based
on these results, our 1-D LHD is sufficient to simulate the head-on GIMs. 
 
The results of three sets of simulations are summarized in Table
\ref{tab:largea} marked as ``L''. They are three sets of models in which the
impactor is a 25 $M_\oplus$ proto-planetary core, a 10 $M_\oplus$
proto-planetary core, and a Saturn-mass gas giant. All of these models are
assumed to be in hydrostatic equilibrium prior to the impact. From virial
theorem \citep{Kippenhahn1990} we find the total energy for a polytropic
equation of state is 

\begin{equation}
W = E_g + E_i = (\zeta -1) E_g/\zeta
\end{equation}
where $E_g$ and $E_i$ are gravitational binding and internal energy, $\zeta= 3
(\gamma-1)$. For $\gamma=5/3$, $E_i = - E_g/2$. At the onset of the
simulation, we impose a burst of thermal energy, $E_a = M_I v_{imp}^2/2$, in
order to take into account the energy released in the vicinity of the core.
Therefore, the total energy becomes $W^\prime = E_a - E_g/2$. We also adopt
the assumption that the cores and embryos are completely merged during the
head-on impacts. This assumption is supported by the results of SPH models
Sa, Sb, and Sc.

For a impactor with mass of $M_I = 25 M_{\oplus}$, a parabolic collision
(model La) leads to a merger without any loss of the gaseous envelope. The gas
giant's photospheric radius expanded slightly. Analogous dynamical parameters
are imposed for a 10 $M_\oplus$ impactor. In this case, the gaseous envelope
is heated by the impact energy and the interior temperature is increased by a
order of magnitude. However, the gravity of the gas giant is also increased
due to the accretion of a 10 $M_\oplus$ proto-planetary core. Therefore, the
expansion of the envelope due to the impact heating is suppressed and the gas
giant contracted a little bit for this model. In Model Lb, the asymptotic
radius after the impact is in fact slightly smaller than its initial radius
as a consequence of the increase in the core mass.  
In the parabolic model Lc, the
merger product produces a much larger photospheric radius than the gas giants'
initial size. Nevertheless, the entire envelope is retained due to a
comparable increase in the merger's thermal and gravitational binding energy.

\subsection{Mixing of cores and envelope material} 
\label{sec:mixingcore}

In the discussion above and in \S\ref{sec:sph}, we showed with SPH
that the impactors with masses of 10 $M_\oplus$ and 25 $M_\oplus$ had adequate mass and intruding velocity to reach
the core of the target planet without total disruption. In this subsection, we
consider the mixing between the two cores and that between the core and the
envelope. The efficiency of mixing determines the evolution of the merger's
internal structure and mass-radius relation.

Massive impactors have $M_{I,c}$'s which are significant fractions of or
comparable to the $M_{T,c}$. The dissipated energy $E_a$ is comparable to the
gravitational binding energy of the target gas giant. If this energy is
efficiently converted to thermal energy throughout the merged core,
temperature in the core would increase by

\begin{equation}
\Delta T_{\rm core} \simeq (\gamma-1) (\mu/ R_g) (E_a/M_{f,c}) \sim 
(v_{\rm imp}/v_{\rm esp})^2 T_i (M_{i, c}/M_{f,c})
\label{eq:tcoretot}
\end{equation} 
where $T_i$ is the core temperature prior to the impact. 

In the models La, $\Delta T_{\rm core} > 5\times 10^4$ K which is most likely
to be larger than that ($T_{\rm evap}$) for heavy elements to fully evaporate
into a gas phase. In this limit, the vapors of heavy elements would thoroughly
mix with the nearby hydrogen gas in the envelope. In the SPH models where the
core and the embryos are approximated as fluids with Tillotson
\citep{Tillotson1962} equation of state for iron and basalt.

At lower temperature, the mixture of composite elements undergoes phase
separation. Although the phase separation diagram is well established for H-He
mixture \citep{Stevenson1976}, its analogs for H-Fe or H-Basalt mixture are
not available. Nonetheless, we expect the heavy elements in the gas phase to
separate from hydrogen into a two-phase medium as the temperature decreases
below a critical value $T_{2p}$. During the phase separation, there is an
exchange between Gibb's free energy and the internal energy
$E_i$ \citep{Stevenson1976}.

After phase separation, density contrast between the two phase increases with
additional reduction in temperature. If the heavy elements can assemble liquid
drops centered around the nucleation sites and grow to sufficiently large
sizes, they would precipitate and settle to the core (Nucleation is a poorly
understood critical phenomenon and it can proceed over a considerable time
scale). Further reduction in planet's internal temperature below some
threshold $T_{\rm cond}$ would lead to the condensation of the heavy elements
\citep{Hubbard1984}.

\subsection{Core growth after impactor's disintegration in the envelope}

We consider that for less massive ($M_I \lesssim$ a few $M_\oplus$) impactors,
especially for oblique collisions, the impactors may disintegrate before
reaching the core. The distribution of heavy elemental deposition $d M_{\rm
dep} / dr$ depends on the impact angle, internal density inside the gas giant,
and the mass of the impactor. Without the loss of generality, we adopt a
simple algebraic expression such that the rate of density added at each radius
is

\begin{equation}
{\partial \rho_{\rm dep}(r) \over \partial t} 
= {\rho_{\rm nor} \over \tau_{\rm dep}}
\left( 1 - \left( {r - r_{\rm dep} \over \Delta r_{\rm dep} }\right)
\right)^\eta {\rm exp} (- {t - t_{\rm imp} \over \tau_{\rm dep}})
\label{eq:rhobreak}
\end{equation}
where $t_{\rm imp}$ and $\tau_{\rm dep}$ represent the epoch of the impact and
the duration of disintegration, $r_{\rm dep}$ and $\Delta r_{\rm rep}$
correspond to the main and spread in the location of disintegration, $\eta$ is
a power index which describes the confinement of disintegration, and
$\rho_{\rm nor}$ is a normalization factor for matching the impact's total
mass. The actual value of $r_{\rm dep}$ is determined by the disintegration
condition, which may have a wide dispersion due to different impact
trajectories and collisional speeds even for the embryos with similar masses.

After the disintegration is completed, $d M_{\rm dep} / dr = 4 \pi \int r^2
(\partial \rho_{\rm dep}(r)/ \partial t) dt$. The associated energy
dissipation rate is

\begin{equation}
{\partial^2 E_{\rm dep} (r) \over \partial r \partial t} = {G M_p (r)
\over r} {\partial^2 M_{\rm dep} \over \partial r \partial t}
\label{eq:enedep}
\end{equation}
and change in the local temperature is 
\begin{equation}
\Delta T_{\rm dep} \simeq {\gamma-1 \over R_g} \mu^\prime {G M_p (r)
\over ( M_{\rm dep}/dr + 4 \pi \rho r^2) r} {d M_{\rm dep} \over dr}
\end{equation} 
where 
\begin{equation}
\mu^\prime = (\mu_{\rm dep} (d M_{\rm dep}/d r) + \mu_{\rm env} 4 \pi
\rho r^2 )/ ( M_{\rm dep}/dr + 4 \pi \rho r^2)
\label{eq:mudep}
\end{equation}
is the modified molecular weight.

Depending on the model parameters, the magnitude of $\Delta T$ may not be
adequate for the new temperature to reach $T_{2 p}$, in which case, the heavy
elements are retained their own identifies in the form of emulsions. Around
the nucleation site, heavy elements congregate and form droplets. Due to the
buoyancy effect, these droplets (with density $\rho_d$ greater than the local
gas density $\rho_g$) have a tendency to precipitate. In an convective region
where the local convection speed is $v_{\rm conv}$, droplets with sizes $s$
larger than a critical size

\begin{equation}
s_{\rm settle} \sim \rho_g v_{\rm conv}^2 / \rho_d g
\end{equation}
(where $g$ is the local gravity), would sediment with a speed
\begin{equation}
v_{\rm settle} \sim (s/s_{\rm settle})^{1/2} v_{\rm conv} \sim (s g
\rho_d/\rho_g)^{1/2}.
\label{eq:vterminal}
\end{equation}
Smaller droplets may also be dragged by the convective eddies, diffuse to
and become part of the core. 

The actual size distribution of the droplets may depend on not only how the
impactors break up, but also how their fragments subsequently interact with
the ambient gas in the envelope. Equation (\ref{eq:rhobreak}) is based on the
assumption that the impactor is completely disintegrated down to the molecular
scales which then form liquid droplets around the nucleation sites. These
liquid droplets coagulate as they collide through dispersive motion or
differential settling. They also fragment due to both collisional
fragmentation as well as Kelvin-Helmholtz and Rayleigh-Taylor instabilities at
their interfaces with the envelope gas. For sufficiently small droplets, these
instabilities are stabilized by their surface tension. Sublimation (from
liquid to gas phase), condensation (from liquid to condensed state), chemical
reaction, and sedimentation can also contribute to the size distribution.

Despite all these uncertainties, we consider a general power-law size
distribution in which $dN/ds \propto s^\lambda$. For solid particles in a gas
free environment, collisional equilibrium is establish with $\lambda \sim
-3.5$ \citep{Dohnanyi1969, Williams1994, Farinella1996,
Tanaka1996, Kenyon2002, Weidenschilling2004}.

The SPH models presented in \S\ref{sec:sph} show differentiation occurs
rapidly. This evolution is possible if the heavy elements can concentrate
into large coherent clouds. However, these results may be an artefact of the
SPH scheme which uses a separate set of independent ``fluid particles'' to
represent the hydrogen gas. This scheme does not take into account the Gibbs
free energy associated with the mixing of two species. On the scale of the
individual fluid elements, they are not subject to additional fragmentation.
Since this scale is $ > > s_{\rm settle}$, sedimentation speed becomes a
significant fraction of $v_{\rm esc}$ and the time scale for differentiation
becomes comparable to that for dynamical collapse.

In principle, the sedimentation of the disintegrated debris should be analyzed
with a two fluid approach. However, if the fragments are sufficiently small to
be well coupled to the gas, the flow can treat the mixture as one gas with a
modified molecular weight $\mu^\prime$. With equations (\ref{eq:rhobreak}),
(\ref{eq:enedep}), and (\ref{eq:mudep}), we include the effect of energy
deposition, mass loading, and molecular weight increases in the LHD models.

In order to simulate an event in which an embryo disintegrates in the planet's
envelope, we introduce a new model Lb1. All the initial and boundary conditions
in this model are identical to those of model Lb, except we set $r_{\rm dep} =
2r_{core}$, $\Delta r_{\rm dep} = 0.3 r_{\rm dep}$, and $\tau_{\rm dep} = 10^6
s$ in equation (\ref{eq:rhobreak}). The normalization factor $\rho_{normal}$ is
adjusted so that the total mass of the impactor remains to be 10 $M_\oplus$.
The associated $E_a = 2.4 \times 10^{41}$ ergs which is similar to that ($2.22
\times 10^{41}$ ergs) for model Lb.

In model Lb1, debris of the impactor is deposited well outside the target gas
giant's core. Over a few dynamical time scales, a quasi hydrostatic equilibrium
is established in the envelope and the flow speed everywhere inside the
photosphere decreases below 0.1 $c_s$. Thereafter planet's interior structure
evolve through thermal adjustments as heat is transferred from its interior and
lost near its photosphere. The efficiency of heat transfer is determined by not
only radiative diffusion but also thermal convection.

According to the standard Schwarzschild criterion, the chemically homogeneous
planetary envelope is generally unstable against convection. However, heavy
element and energy deposition after a major impact modify the local thermal and
compositional distribution. According to the Ledoux criterion, the flows can be
stabilized by the molecular weight gradient if

\begin{equation}
\nabla_{T} < \nabla_{ad} + (\varphi/\delta) \nabla_\mu
\label{eq:ledoux}
\end{equation}
where the radiative $\nabla_{T} = (P/T) (dT/dP)= (3 \kappa F/4 ac T^2) (P/g)$,
the adiabatic $\nabla_{ad} = (\gamma-1)/\gamma \sim 0.4$, chemical $\nabla_\mu
= (d {\rm ln} \mu / d {\rm ln} P)_s$. The subscript $s$ refers to the
surrounding medium. For an ideal gas, both $\varphi (\equiv d {\rm ln} \rho / d
{\rm ln} \mu$) and $\delta (\equiv - d {\rm ln} \rho / d {\rm ln} T)$ equal to
unity \citep{Kippenhahn1990}.

\begin{figure}
\begin{center}
\epsscale{1}\plotone{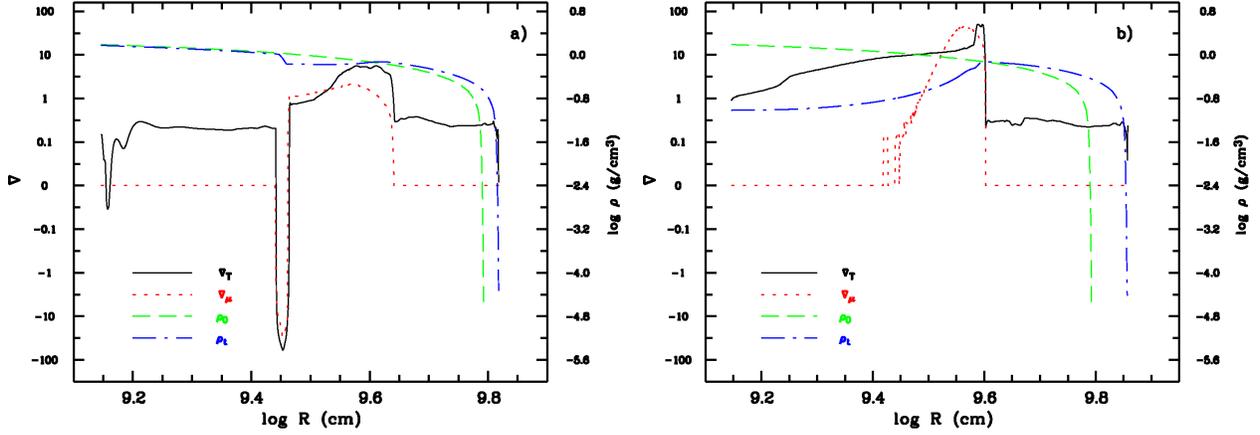}
\end{center}
\caption{Gradients for temperature and molecular weight distributions for
models Lb1 (on the left panel) and Lb2 (on the right panel) after the models
have reached quasi-static equilibrium. The black solid line shows the
temperature gradient. The red dotted line corresponds to the molecular weight
gradient. The density profiles for both models are overplotted on the same
plots. The green dashed line represents the density profile of the initial
model. The blue dash-dotted line shows the density profile at the same epoch as
when those gradients are plotted.} 
\label{fig:thermal} 
\end{figure}

The initial density distribution for model Lb1 is plotted on the left panel of
Figure\ref{fig:thermal}. Along with the density distribution, overplotted are
the $\nabla_T$, and $\nabla_\mu$ distribution shortly after a quasi hydrostatic
equilibrium is reached. At this epoch, there are three interesting regions. 

a) Outside $r=3.3 \times 10^9$ cm, the envelope is convectively unstable
because both Schwarzschild and Ledoux criterion for stability are not
satisfied. The impulsive heat deposition drives vigorous convection to
transport heat and to mix the heavy elements with the envelope gas.

b) In the region $r=3\times 10^9 - 3.3 \times 10^9$ cm, where the Ledoux criterion for
stability is satisfied but Schwarzschild criterion is not, the envelope is
dynamically stable but vibrationally unstable (or over-stable)
\citep{Kippenhahn1990}. In this limit, double (thermal and chemical) diffusion
leads to layered structure in which diffusive radiative layers are sandwiched
between convective layers which are well mixed and have adiabatic structures.
When averaged over many layers, stratification would probably adjust toward a
state of marginal stability. With an equation of state, $d \rho/\rho = \alpha
(d P /P) - \delta (d T/T) + \psi (d \mu / \mu)$, the Ledoux criterion is
marginally satisfied with

\begin{equation}
\nabla_{\rm tot} = \nabla_{\rm rad} - \nabla_{\rm adiabatic}-
\nabla_\mu (\psi / \delta) \simeq 0.
\end{equation} 

c) Interior to $r= 3 \times 10^9$ cm, the flow is stable according
to both criterion and the initial convective envelope is stabilized by
the heat released and heavy element deposition. 

\begin{figure}
\begin{center}
\epsscale{1}\plotone{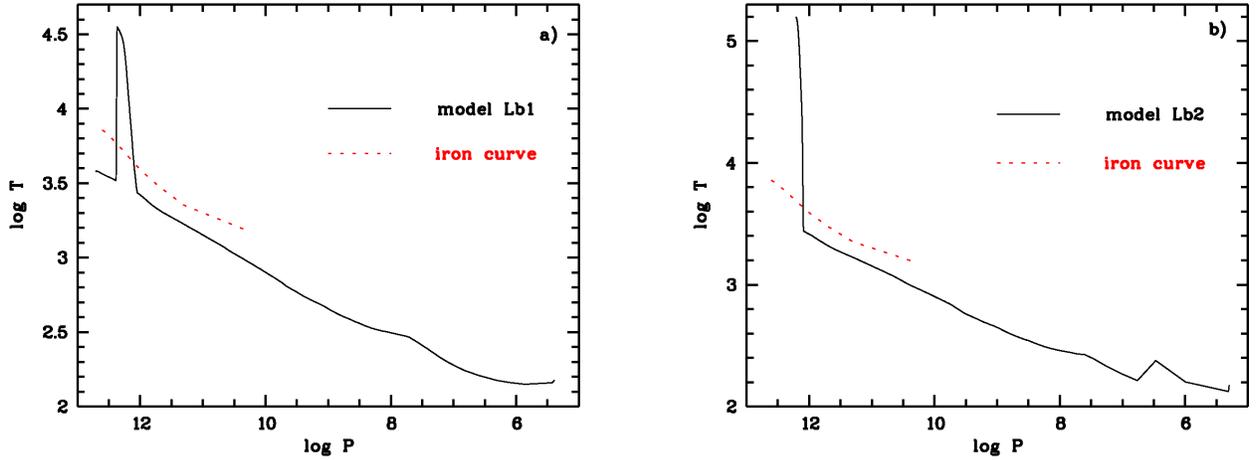}
\end{center}
\caption{P-T diagrams for model Lb1 (on the left panel) and Lb2 (on the right
panel). Overplotted is the melting curve of iron.}
\label{fig:PTdiag}
\end{figure}

In both regions b) and c), convection is suppressed with vanishing $v_{conv}$.
After the impact, the temperature in these regions is well above $T_{cond}$
(see Figure \ref{fig:PTdiag} panel a) and most likely above $T_{2 p}$. Despite the
stabilization of regions b) and c) against convection, heat is still transferred
to the planet's surface through radiative diffusion. On a much longer
Kelvin-Helmholtz cooling time scale ($\sim 10-100$ Myr), phase separation would
lead to the formation of liquid droplets when the envelope's temperature
declines below $T_{2 p}$, provided that the condition allows the conversion of
Gibb's free energy to internal energy. 

Droplets grow through condensation and coagulation. While the convective
stability is maintained, relatively large droplets would sediment towards the
core with a terminal speed $v_{settle} \sim (s g \rho_d/\rho_g)^{1/2} \sim (s /
R_p)^{1/2} v_{esc}$. In this Stoke's range, the sedimentation time scale
($\tau_{settle} \sim (R_p/s)^{1/2} \tau_{dyn}$, where $\tau_{dyn} \sim (R_p^3
/GM_p)^{1/2}$ is the dynamical time scale) is shorter than the age of mature
solar-type star, $\tau_\ast$. 

Drag on the small particles (with $s < s_{mfp}$, where the mean free path
$s_{mfp} \simeq 5 \times 10^{-9} cm / (\rho / 1 g cm^{-3})$ for the ambient
hydrogen gas) is in the Epstein range. In this limit, $ v_{settle} \sim (s g/
c_s) (\rho_d /\rho_g) \sim (s / R_p) v_{esc}$ and $\tau_{settle} \sim (R_p/s)
\tau_{dyn}$ which may be substantially longer than $\tau_\ast$.

The above consideration indicates that the debris deposit of moderate-mass
embryos which disintegrated in the planetary envelopes not only introduces a
metal-enriched layer well outside the core but also locally suppress convection.
These two effects promotes fragments sedimentation and the growth of their
cores.

\subsection{Core erosion after impactor's disruption near the core}

It has long been recognized that compositional stabilization is an important
process in preserving the heavy-element-rich cores against thermal convection
of the gaseous envelopes in isolated gas giants \citep{Guillot2004}. If
turbulence in the convective envelopes can reach the cores, mixing and erosion
would reduce $M_{t,c}$. However, this diffusion leads to a strong negative
molecular weight gradient across the core-envelope interface. This gradient
would quench thermal convective instabilities, suppress the compositional
diffusion rate, and preserve the core-envelope separation. 

In \S \ref{sec:sph} and \S \ref{sec:lhd}, we showed that in nearly head-on
GIMs, sufficiently massive impactors can penetrate deeply into the target gas
giants' envelopes and reach their cores. For these GIMs we show in this
subsection, that the impulsive dissipation of impact energy would strongly
drive thermal convection despite the compositional gradient introduced by the
heavy element injection and diffusion.

In order to demonstrate this possibility, we construct model Lb2, in which,
debris of the impactor is deposited in the proximity of the target gas giant's
core. All the initial and boundary conditions in this model is identical to
those of model Lb, except we set $r_{\rm dep} =r_{\rm core}$, $\Delta r_{\rm
dep} = 0.25 r_{\rm dep}$, and $\tau_{\rm dep} = 1\times 10^6$ s in equation
(\ref{eq:rhobreak}). The normalization factor $\rho_{\rm normal}$ is adjusted
so that the total mass of the impactor remains to be 10 $M_\oplus$. The
associated $E_a = 1.61 \times 10^{41}$ ergs.

Similar to model Lb1, a quasi hydrostatic equilibrium is established in the
envelope and the flow speed everywhere decreases below 0.1 $c_s$ after a few
dynamical time scale. We plot $\rho_g$, $\nabla_T$ and $\nabla_\mu$
distribution at this epoch for model Lb2 on the right panel of Figure
\ref{fig:thermal}. By this stage, a fraction of the debris of the disintegrated
impactors has reached $r=4 \times 10^9$cm. Within this radius, both
Schwarzschild and Ledoux criterion for stability are violated and the heat
release near the core strongly enhances the tendency for thermal convection
throughout the envelope, including the boundary layer region.

The subsequent evolution of the heavy elements depends on whether the
collisions lead to a total phase transition of the core. Let us first consider
the possibility of a hot core with an initial temperature $T_{\rm core} >T_{\rm
cond}$ so that the heavy elements are in a liquid or gas state. Provided the
core-envelope interface is stabilized by an initial compositional gradient in
accordance with the Ledoux criterion, this differentiated structure would be
preserved.

After the impact, vigorous and efficient convection would induce the interior
of the planet to evolve towards a new constant-entropy state. During that
evolution, if a fraction of the dissipated energy $f E_a$ is transported into
the central region, the fractional change in the core would be $\Delta T_{\rm
core} /T_{\rm core} \sim f E_a/ E_i$ (In the limit $E_a > E_i$, $\Delta T_{\rm
core}$ would be given by equation \ref{eq:tcoretot}). The time scale for
thermal adjustment towards this state is approximately 

\begin{equation}
\tau_{\rm ad} \sim f E_a / 4 \pi R_{\rm core}^2 F_{\rm core}
\end{equation}
where $F_{\rm core}$ is the rate of heat transport into the
core (In the limit of high heat conduction rate, $F_{\rm core}$ would
be the convection flux near the core radius $R_{\rm core}$ and it can
be computed using standard mixing length prescription). During this
time, the amount of heat loss from the planetary surface is $\sim 4
\pi R_p ^2 F_{\rm sur}$ where $F_{\rm sur}$ is the planet's intrinsic
heat flux at its surface (Note $F_{\rm sur}$ does not include the
reprocessed stellar irradiation). From energy conservation, we find
$f \sim ( 1 + (F_{\rm sur} / F_{\rm con}) (R_p/R_{\rm core}) ^2$.

With sufficient input energy $E_a$, the core temperature can increase above
$T_{\rm evap}$ so that the heavy elements are evaporated. In the limit of
efficient convection, the envelope's temperature distribution adjusts to
establish constant entropy state. Heavy element would mix with the gas in
regions where the local temperature is above $T_{2 p}$. If this radius extends
well beyond the initial $R_{core}$, a substantial fraction of the core would be
eroded and mixed with the envelope. However, if after the impact, $T_{\rm core}
^\prime = T_{\rm core} + \Delta T_{\rm core} < T_{\rm evap}$ or the region with
$T> T_{2 p}$ is confined to small radius, the core structure would be mostly
preserved even though convection may be able to penetrate through the initial
core-envelope interface.

We now consider the possibility of GIMs onto gas giants with cool ($T_{\rm
core} < T_{\rm cond}$) cores. With sufficiently massive and energetic impactor,
head-on collisions lead to $T_{\rm core} ^\prime > T_{\rm evap}$. The
subsequent evolution is similar to that discussed above, albeit a modest
fraction of internal energy must be used to compensate for the latent heat
during the melting and evaporation of the core. Less energetic impacts may not
be able to evaporate the core, in which case, the temperature distribution of
the core would be determined by the energy equation (\ref{eqn:energy})
including heat transport by conduction. Provided the new $T > T_{\rm cond}$ at
the core-envelope interface, the molten heavy elements would be entrained into
the convective eddies on the growth time scale for Kelvin-Helmholtz
instability. This rate may be limited by the poorly determined material
properties such as surface tension and phase separation. However, if $T > T_{2
p}$ at the core-envelope interface, mixing between the heavy elements and
hydrogen may be much more efficient. Since convection in the envelope extents
to the interface, the mixed gas is redistributed throughout the convective
region with $T > T_{2 p}$.

\section{Summary and Discussions}
\label{sec:summary}

The main objective of this paper is to suggest that giant impacts and merger
process may have led to different core-envelope structure between Jupiter and
Saturn and the super-solar metallicity in their gaseous envelopes.

We have highlighted several avenues of GIMs by embryos onto gas giants. The
supply of these large solid building blocks include: {\it in situ} embryos
during the final phase of gas giant formation, and terrestrial bodies cleared
by sweeping secular resonances and dynamical instabilities. There are also
several processes which can lead to the merger of gas giants, including orbit
crossing triggered by run-away migration, and long-term dynamical instability.

At their present-day orbits, the velocity dispersion of any residual embryos or
planetesimals is less than the gas giants' surface escape speed. The GIMs
between any residual planetesimals, embryos, or hypothetic gas giants and
long-period gas giants are likely to be parabolic and attain limited amount of
energy. This modest velocity dispersion also enlarges gas giants' gravitational
cross section. 

The discussions above indicate that embryos with a mass of several $M_\oplus$
can penetrate through the envelope and reach the core of gas giants through
head-on collisions. If these collisions are sufficiently energetic, they would
lead to significant core erosion. Heavy elements would be well mixed with the
hydrogen in the envelope down to a molecular level. The average density and
metallicity of the gas in the envelope would be enhanced. Although a relatively
large amount of kinetic energy is dissipated into internal energy, the
heat released from this type of GIM is efficiently transported to the planet's
surface and radiated away because the envelope is fully convective.

Low-mass impactors ($M_I\ll M_{\oplus}$ are likely to disintegrate in
gas giants' envelope. Energy deposition in 
the planet's envelope suppresses convection interior to $r_{dep}$. Heavy
element deposition also increases the opacity. After the planet has adjusted to
a new hydrostatic equilibrium, the impact dissipation energy is more
effectively retained by the reduced heat flux through the envelope. Suppression
of convection also leads to sedimentation of modest to large size embryos. In
the limit that a collisional equilibrium can be established with most of the
debris' mass being in the forms of large particles, the core attains, from the
post-impact differentiation, a mass substantially larger than the critical mass
(a few $M_\oplus$) for the onset of efficient gas accretion \citep{Pollack1996,
Ida2004}.

We suggest these two limiting cases introduce diversity in the core-envelope
structure. They may also lead to different cooling rates during the subsequent
thermal evolution. For the core-impact case, the dredge-up of heavy elements
does not suppress convection. During the cooling process, phase separation and
phase transition lead to changes in the Gibbs free energy which may lead to the
release of additional sources of internal energy to reduce the planet's
contraction rate. For the envelope-disintegrate case, small grains or droplets
are the dominant sources of opacity \citep{Depater2001} and their suspension
or depletion determines the contraction rate of the gas giant. Analysis for the
long term thermal evolution of planets will be considered and presented
elsewhere.

We suggest that Jupiter may have been struck by an energetic head-on GIM which
penetrated deeply into its envelope and lead to its core erosion. With a larger
($>10 M_\oplus$) initial core, it would also partly resolve a long standing
issue on how Jupiter might acquire its massive gaseous envelope with
such a small core on the multi-Myr time scale over which their natal
solar nebula is depleted.

In contrast, Saturn may have acquired its present big core through GIMs with
less energetic and smaller residual embryos. A temporary stabilization of
Saturn's envelope against convection may have also prolonged its contraction
time scale and enabled the precipitation of helium dropout in its envelope. We
note that both processes lead to the enrichment of metallicity in the envelopes
of Jupiter and Saturn.

GIMs are most likely to occur within the first few Myr of their life span.
Energy dissipation from such events is likely to increase the intrinsic
luminosity and radius of these planets. Direct imaging and rare transit
observations may provide observational verifications that the internal
structure of typical gas giants may have been modified by GIMs. We shall
discuss the GIM's influence on the mass-radius-luminosity of the close-in
planets in Paper II.

We have shown possible mechanisms for the dredge up of heavy elements from the
core and for the promotion of core growth. We have not considered here the
long-term sustainability of the stir-up elemental distribution. The follow up
evolution of the impacted gas giant's radius depends sensitively on their
cooling efficiency and elemental segregation. Although there are models for
contraction and differentiation of initially homogeneous gas
giants \citep{Bodenheimer2003, Baraffe2008}, the long-term post GIM
evolution still needs to be investigated. 

\acknowledgements We thank Drs P. Bodenheimer, P. Garaud, T. Guillot, and
G. Laughlin for useful conversation. Part of the calculation is carried out on the SGI Altix 330 system at
the Department of Astronomy, Peking University. This work is supported by NASA
NNX07A-L13G, NNX07AI88G, NNX08AL41G, NNX08AM84G, and NSF(AST-0908807). 
CBA's involvement was supported by NASA (NNG05G1496).


\begin{thebibliography}{}

\bibitem[\protect\citeauthoryear{{Agnor} \& {Asphaug}}{{Agnor} \& {Asphaug}}{2004}]{Agnor2004}
{Agnor} C., {Asphaug} E., 2004, \apjl, 613, L157

\bibitem[\protect\citeauthoryear{{Alexander} \& {Ferguson}} {{Alexander} \&
{Ferguson}}{1994}]{Alexander1994} {Alexander} D. R., {Ferguson} J. W., 1994, \apj,
437, 879

\bibitem[Anic et 
al.(2007)]{Anic_etal_2007} Anic, A., Alibert, Y., \& Benz, W.\ 2007, \aap, 466, 717 


\bibitem[\protect\citeauthoryear{{Asphaug} et al.}{{Asphaug} et al.}{1998}]
{Asphaug1998} {Asphaug} E., {Ostro} S. J., {Hudson} R. S., {Scheeres} D. J.,
{Benz} W., 1998, \nat, 393, 437


\bibitem[Asphaug et al.(2006)]{Asphaug_etal_2006} Asphaug, E., Agnor, 
C.~B., \& Williams, Q.\ 2006, \nat, 439, 155 


\bibitem[\protect\citeauthoryear{{Balbus} \& {Hawley}}{{Balbus} \&
{Hawley}}{1991}]{Balbus1991} {Balbus} S. A., {Hawley} J. F., 1991, \apj,
376, 214

\bibitem[\protect\citeauthoryear{{Baraffe} et al.}{{Baraffe} et al.}{2008}]
{Baraffe2008} {Baraffe} I., {Chabrier} G., Barman T., 2008, A\&A, 482, 315


\bibitem[\protect\citeauthoryear{{Benz}}{{Benz}}{1990}]
{Benz1990} {Benz} W., 1990, in Numerical Modelling of Nonlinear Stellar
Pulsations Problems and Prospects, ed. {Buchler} J. R., (NATO ASI Ser. C, 302;
Dordrecht: Kluwer), 269

\bibitem[Benz et al.(2007)]{Benz_etal_2007} Benz, W., Anic, A., 
Horner, J., \& Whitby, J.~A.\ 2007, Space Science Reviews, 132, 189 

\bibitem[\protect\citeauthoryear{{Benz} \& {Asphaug}}{{Benz} \&
{Asphaug}}{1999}]{Benz1999} {Benz} W., {Asphaug} E., 1999,
Icarus, 142, 5

\bibitem[\protect\citeauthoryear{{Benz} \& {Hills}}{{Benz} \&
{Hills}}{1987}]{Benz1987} {Benz} W., {Hills} J. G., 1987, \apj, 323, 614

\bibitem[\protect\citeauthoryear{{Benz} et al.}{{Benz} et al.}{1986}]
{Benz1986} {Benz} W., {Slattery} W. L., {Cameron} A. G. W., 1986,
Icarus, 66, 515


\bibitem[Benz et al.(1988)]{Benz_etal_1988} Benz, W., Slattery, W.~L., 
\& Cameron, A.~G.~W.\ 1988, Icarus, 74, 516 




\bibitem[\protect\citeauthoryear{{Bryden} et al.}{{Bryden} et al.}{2000}]{Bryden2000a}
{Bryden} G., R{\' o}{\` z}yczka M., {Lin} D. N. C., {Bodenheimer} P., 2000a, ApJ, 540, 1091

\bibitem[\protect\citeauthoryear{{Bryden} et al.}{{Bryden} et al.}{2000}]{Bryden2000b}
{Bryden} G., {Lin} D. N. C., {Ida} S., 2000b, ApJ, 544, 481

\bibitem[\protect\citeauthoryear{{Bodenheimer} et~al.}{{Bodenheimer}
et~al.}{2000}]{Bodenheimer2000} {Bodenheimer} P., {Hubickyj} O.,
{Lissauer} J. J., 2000, Icarus, 143, 2

\bibitem[\protect\citeauthoryear{{Bodenheimer} et~al.}{{Bodenheimer}
et~al.}{2001}]{Bodenheimer2001} {Bodenheimer} P., {Lin} D. N. C.,
{Mardling}, R. A., 2001, \apj, 124, 62

\bibitem[\protect\citeauthoryear{{Bodenheimer} et~al.}{{Bodenheimer}
et~al.}{2003}]{Bodenheimer2003} {Bodenheimer} P., {Laughlin} G.,
{Lin} D. N. C., 2003, \apj, 592, 555

\bibitem[\protect\citeauthoryear{{Bodenheimer} \& {Pollack}}{{Bodenheimer} \&
{Pollack}}{1986}]{Bodenheimer1986} {Bodenheimer} P., {Pollack} J. B., 1986,
Icarus, 67, 391

\bibitem[\protect\citeauthoryear{Boss}{Boss}{1997}]{Boss1997} {Boss} A. P.,
1997, Science, 276, 1836 

\bibitem[\protect\citeauthoryear{{Burkert} et al.}{{Burkert} et al.}{2005}]
{Burkert2005} {Burkert} A., {Lin} D. N. C., {Bodenheimer} P. H., {Jones} C. A.,
{Yorke} H. W., 2005, \apj, 618, 512

\bibitem[\protect\citeauthoryear{{Burrows} et al.}{{Burrows} et al.}{2000}]
{Burrows2000} {Burrows} A., {Guillot} T., {Hubbard} W. B.,{Marley} M. S., 
{Saumon} D., {Lunine} J. I., {Sudarsky} D., 2000, \apj, 534, 97

\bibitem[\protect\citeauthoryear{{Burrows} et al.}{{Burrows} et al.}{2004}]
{Burrows2004} {Burrows} A., {Hubeny} I., {Hubbard} W. B., {Sudarsky} D.,
{Fortney} J. J., 2004, \apjl, 610, L53


\bibitem[Cameron \& Ward(1976)]{Cameron_&_Ward_1976} Cameron, A.~G.~W., \&
Ward, W.~R.\ 1976, Lunar and Planetary Institute Science Conference Abstracts,
7, 120 


\bibitem[Canup \& Asphaug(2001)]{Canup_&_Asphaug_2001} Canup, R.~M., \&
Asphaug, E.\ 2001, \nat, 412, 708 


\bibitem[\protect\citeauthoryear{{Chatterjee} et~al.}{{Chatterjee}
et~al.}{2008}]{Chatterjee2008} {Chatterjee} S., {Ford} E. B., {Matsumura} S.,
{Rasio} F. A., 2008, \apj, 686, 580

\bibitem[\protect\citeauthoryear{{Cumming} et~al.}{{Cumming}
et~al.}{2008}]{Cumming2008} {Cumming} A., {Butler} R. P., {Marcy} G. W.,
{Vogt} S. S., {Wright} J. T., {Fischer} D. A., 2008, \pasp, 531, 120

\bibitem[\protect\citeauthoryear{{Cuzzi} \& {Zahnle}}{{Cuzzi} \&
{Zahnle}}{2004}]{CuzziZahnle2004} {Cuzzi} J.~N., {Zahnle} K.~J., 2004,
\apj, 614, 490

\bibitem[\protect\citeauthoryear{{de Pater} \& {Lissauer}}{{de Pater} \&
{Lissauer}}{2001}]{Depater2001} {de Pater} I., {Lissauer} J.~J., 2001,
Planetary Sciences, (UK: Cambridge University Press)


\bibitem[\protect\citeauthoryear{{Dobbs-Dixon} et~al.}{{Dobbs-Dixon}
et~al.}{2007}]{Dobbs-Dixon2007} {Dobbs-Dixon} I., {Li}, S.-L., {Lin} D. N. C.,
2007, \apj, 660, 791

\bibitem[\protect\citeauthoryear{{Dodson-Robinson} et~al.}{{Dodson-Robinson}
et~al.}{2008}]{DodsonRobinson2008} {Dodson-Robinson} S. E., {Willacy} K.,
{Bodenheimer} P., {Turner} N. J., {Beichman} C. A., 2008, astro-ph/0806.3788

\bibitem[\protect\citeauthoryear{{Dohnanyi}}{{Dohnanyi}}{1969}]{Dohnanyi1969}
{Dohnanyi} J.~W., 1969, \jgr, 74, 2531

\bibitem[\protect\citeauthoryear{{Duncan} et~al.}{{Duncan}
et~al.}{1987}]{Duncan1987} {Duncan} M., {Quinn} T., {Tremaine} S., 1987,
\aj, 94, 1330

\bibitem[\protect\citeauthoryear{{Farinella} \& {Davis}}{{Farinella} \&
{Davis}}{1996}]{Farinella1996} {Farinella} P., {Davis} D.~R., 1996,
Science, 273, 938

\bibitem[\protect\citeauthoryear{{Gammie}}{{Gammie}}{1996}]{Gammie1996}
{Gammie} C.~F, 1996, \apj, 457, 355

\bibitem[\protect\citeauthoryear{{Garaud} et~al.}{{Garaud} et~al.}{2004}]{Garaud2004}
{Garaud} P., {Barri{\`e}re-Fouchet} L., {Lin} D. N. C., 2004, \apj, 603, 292

\bibitem[\protect\citeauthoryear{{Garaud} \& {Lin}}{{Garaud} \&
{Lin}}{2007}]{Garaud2007} {Garaud} P., {Lin} D. N. C., 2007, \apj, 654, 666

\bibitem[\protect\citeauthoryear{Gillon et al.}{Gillon et al.}{2007}]{Gillon2007}
{Gillon} M., {Pont} F., {Moutou} C., {Santos} N. C., {Bouchy} F., {Hartman} J. D.,
{Mayor} M., {Melo} C., {Queloz} D., {Udry} S., {Magain} P., 2007, A\&A, 466, 743

\bibitem[\protect\citeauthoryear{{Goldreich} \& {Tremaine}}{{Goldreich} \& {Tremaine}}{1980}]{Goldreich1980}
{Goldreich} P., {Tremaine} S., 1980, ApJ, 241, 425


\bibitem[\protect\citeauthoryear{{Gu} et~al.}{{Gu} et~al.}{2003}]{Gu2003}
{Gu} P. G., {Lin} D. N. C., {Bodenheimer} P. H., 2003, \apj, 588, 509 

\bibitem[\protect\citeauthoryear{{Gu} et~al.}{{Gu} et~al.}{2004}]{Gu2004}
{Gu} P. G., {Bodenheimer} P. H., {Lin} D. N. C., 2004, \apj, 608, 1076

\bibitem[\protect\citeauthoryear{{Guillot} et~al.}{{Guillot}
et~al.}{2004}]{Guillot2004} {Guillot} T., {Stevenson} D. J., {Hubbard} W. B.,
{Saumon} D., 2004, in Jupiter. The Planet, Satellites, and Magnetosphere,
ed. {Bagenal} F., {Dowling} T. E., {McKinnon} W. B.,
(Cambridge: Cambridge Univ. Press), 35

\bibitem[Hahn \& Malhotra(1999)]{Hahn1999} Hahn, J.~M., \&
Malhotra, R., 1999, \aj, 117, 3041

\bibitem[\protect\citeauthoryear{{Hayashi} et~al.}{{Hayashi}
et~al.}{1985}]{Hayashi1985} {Hayashi} C., {Nakazawa} K., {Nakagawa} Y.,
1985, in Protostars and Planets II, ed. {Black} D. C., {Matthews} M. S.,
(Tucson: Univ. Arizona Press), 1100



\bibitem[Hartmann \& Davis(1975)]{Hartmann_&_Davis_1975} Hartmann, W.~K., \&
Davis, D.~R., 1975, Icarus, 24, 504 


\bibitem[\protect\citeauthoryear{{Helled} et~al.}{{Helled}
et~al.}{2008}]{Helled2008} {Helled} R., {Podolak} M., {Kovetz} A.,
2008, Icarus, 195, 863 

\bibitem[\protect\citeauthoryear{{Hubbard}}{{Hubbard}}{1984}]{Hubbard1984}
{Hubbard} W. B., 1984, Planetary interiors (New York, Van Nostrand Reinhold)

\bibitem[\protect\citeauthoryear{{Hubickyj} et~al.}{{Hubickyj}
et~al.}{2005}]{Hubickyj2005} {Hubickyj} O., {Bodenheimer} P., {Lissauer} J. J.,
2005, Icarus, 179, 415

\bibitem[\protect\citeauthoryear{{Ida} \& {Lin}}{{Ida} \&
{Lin}}{2004}]{Ida2004} {Ida} S., {Lin} D. N. C., 2004, \apj, 604, 388

\bibitem[\protect\citeauthoryear{{Ida} \& {Lin}}{{Ida} \&
{Lin}}{2008a}]{Ida2008a} {Ida} S., {Lin} D. N. C., 2008, \apj, 673, 487

\bibitem[\protect\citeauthoryear{{Ida} \& {Lin}}{{Ida} \&
{Lin}}{2008b}]{Ida2008b} {Ida} S., {Lin} D. N. C., 2008, \apj, 685, 584

\bibitem[\protect\citeauthoryear{{Ikoma} et~al.}{{Ikoma} et~al.}{2006}]{Ikoma2006}
{Ikoma} M., {Guillot} T., {Genda} H., {Tanigawa} T., {Ida} S., 2006,
\apj, 650, 1150

\bibitem[\protect\citeauthoryear{{Inaba} \& {Ikoma}}{{Inaba} \&
{Ikoma}}{2003}]{Inaba2003} {Inaba} S., {Ikoma} M., 2003, A\&A, 410, 711

\bibitem[\protect\citeauthoryear{{Kenyon}}{{Kenyon}}{2002}]{Kenyon2002}
{Kenyon} S.~J., 2002, \pasp, 114, 265

\bibitem[\protect\citeauthoryear{{Kippenhahn} \& {Weigert}}{{Kippenhahn} \&
{Weigert}}{1990}]{Kippenhahn1990} {Kippenhahn} R., {Weigert} A., 1990,
Stellar Structure and Evolution (Berlin: Springer)

\bibitem[\protect\citeauthoryear{{Kretke} \& {Lin}}{{Kretke} \&
{Lin}}{2007}]{Kretke2007} {Kretke} K. A., {Lin} D. N. C., 2007, \apjl, 664, L55

\bibitem[\protect\citeauthoryear{{Kretke} et al.}{{Kretke} et al.}{2008}]{Kretke2008}
 {Kretke} K. A., {Lin} D. N. C., {Garaud} P., {Turner} N. J., 2008, astro-ph/0806.1521

\bibitem[\protect\citeauthoryear{{Kokubo} \& {Ida}}{{Kokubo} \&
{Ida}}{2002}]{Kokubo2002} {Kokubo} E., {Ida} S., 2002, \apj, 581, 666

\bibitem[\protect\citeauthoryear{K{\" o}nigl}{K{\" o}nigl}{1991}]{Konigl1991} {K{\" o}nigl} A., 1991,
ApJ, 370, L39

\bibitem[\protect\citeauthoryear{{Korycansky} \& {Zahnle}} {{Korycansky} \&
{Zahnle}}{2005}]{Korycansky2005} {Korycansky} D. G., {Zahnle} K. J., 2005,
\planss, 53, 695

\bibitem[\protect\citeauthoryear{{Lecar} et~al.}{{Lecar} et~al.}{2006}]{Lecar2006}
{Lecar} M., {Podolak} M., {Sasselov} D., {Chiang} E., 2006, \apj, 640, 1115

\bibitem[\protect\citeauthoryear{{Lee} \& {Peale}}{{Lee} \&
{Peale}}{2002}]{Lee2002} {Lee} M. H., {Peale} S. J., 2002, \apj, 567, 596

\bibitem[\protect\citeauthoryear{{Lesur} \& {Ogilvie}}{{Lesur} \&
{Ogilvie}}{2010}]{LesurOgilvie2010} {Lesur} G., {Ogilvie} G.~I , 2010,
MNRAS, 404, L64

\bibitem[\protect\citeauthoryear{{Li} et al.}{{Li} et al.}{2008}]{Li2008}
{Li} S.-L., {Lin} D. N. C., {Liu} X.-W., 2008, \apj, 685, 1210

\bibitem[\protect\citeauthoryear{{Li} et al.}{{Li} et al.}{2010}]{Li2009b}
{Li} S.-L., {Agnor} C., {Lin} D. N. C., 2010, in preparation 

\bibitem[\protect\citeauthoryear{{Lin} et~al.}{{Lin} et~al.}{1996}]{Lin1996}
{Lin} D. N. C., {Bodenheimer} P., {Richardson} D., 1996, Nature, 380, 606

\bibitem[\protect\citeauthoryear{{Lin} \& {Ida}}{{Lin} \&
{Ida}}{1997}]{Lin1997} {Lin} D. N. C., {Ida} S., 1997, \apj, 477, 781

\bibitem[\protect\citeauthoryear{{Lin} \& {Papaloizou}}{{Lin} \&
{Papaloizou}}{1979}]{Lin1979} {Lin} D. N. C., {Papaloizou} J. C. B., 1979,
MNRAS, 186, 799 (or 188, 191)

\bibitem[\protect\citeauthoryear{{Lin} \& {Papaloizou}}{{Lin} \&
{Papaloizou}}{1980}]{LinPap1980} {Lin} D. N. C., {Papaloizou} J. C. B., 1980,
MNRAS, 191, 37

\bibitem[\protect\citeauthoryear{{Lin} \& {Papaloizou}}{{Lin} \&
{Papaloizou}}{1986}]{Lin1986} {Lin} D. N. C., {Papaloizou} J. C. B., 1986,
\apj, 309, 846

\bibitem[\protect\citeauthoryear{{Lin} \& {Papaloizou}}{{Lin} \& {Papaloizou}}{1993}]{Lin1993}
{Lin} D. N. C., {Papaloizou} J. C. B., 1993, in Protostars and Planets III, ed. E. H. Levy
\& J. I. Lunine (Tucson: Univ. Arizona Press), 749

\bibitem[\protect\citeauthoryear{{Lin} et~al.}{{Lin} et~al.}{2009}]{Lin2009}
{Lin} D. N. C., {Zhou} J. L., {Wang} S., {Kretke} K., 2009, in preparation

\bibitem[\protect\citeauthoryear{{Lissauer}}{{Lissauer}}{1987}]{Lissauer1987} 
{Lissauer} J. J., 1987, Icarus, 69, 249

\bibitem[\protect\citeauthoryear{{Love} \& {Ahrens}}{{Love} \&
{Ahrens}}{1996}]{Love1996} {Love} S. G., {Ahrens} T. J, 1996, Icarus,
124, 141

\bibitem[\protect\citeauthoryear{{Mandell} \& {Sigurdsson}}{{Mandell} \&
{Sigurdsson}}{2003}]{Mandell2003} {Mandell} A. M., {Sigurdsson} S., 2003,
\apjl, 599, L111

\bibitem[\protect\citeauthoryear{{Mandushev} et~al.}{{Mandushev}
et~al.}{2007}]{Mandushev2007} {Mandushev} G., {O'Donovan} F. T., {Charbonneau} D., et~al.
2007, \apjl, 667, L195

\bibitem[\protect\citeauthoryear{{Mardling} \& {Lin}}{{Mardling} \& {Lin}}{2004}]{Mardling2004}
{Mardling} R. A., {Lin} D. N. C., 2004, \apj, 614, 955

\bibitem[\protect\citeauthoryear{{Mardling}}{{Mardling}}{2007}]{Mardling2007}
{Mardling} R. A., 2007, MNRAS, 382, 1768

\bibitem[\protect\citeauthoryear{{Masset}}{{Masset}}{2001}]{Masset2001}
{Masset} F. S., 2001, \apj, 558, 453

\bibitem[\protect\citeauthoryear{{Masset}}{{Masset}}{2008}]{Masset2008}
{Masset} F. S., 2008, in IAU Symposium 249, Exoplanets: Detection, Formation
and Dynamics, ed. Y.-S. Sun, S. Ferraz-Mello, \& J.-L. Zhou, 331

\bibitem[\protect\citeauthoryear{{Masset} et~al.}{{Masset}
et~al.}{2006}]{Masset2006} {Masset} F. S., {D'Angelo} G., {Kley} W.,
2006, \apj, 652, 730

\bibitem[\protect\citeauthoryear{{Mayor} et~al.}{{Mayor}
et~al.}{2009}]{Mayor2009} {Mayor} M., {Udry} S., {Lovis} C., {Pepe} F.,
{Queloz} D., {Benz} W., {Bertaux} J. L., {Bouchy} F., {Mordasini} C.,
{Segransan} D., 2009, \aa, 493, 639

\bibitem[\protect\citeauthoryear{{Melosh}}{{Melosh}}{1989}]{Melosh1989}
{Melosh} H. J., 1989, in Impact Cratering (New York: Oxford Univ. Press), 45

\bibitem[\protect\citeauthoryear{{Militzer} et~al.}{{Militzer}
et~al.}{2008}]{Militzer2008}
{Militzer} B., {Hubbard} W. B., {Vorberger} J.,	{Tamblyn} I.,
{Bonev} S. A., 2008, \apjl, 688, L45

\bibitem[\protect\citeauthoryear{{Morbidelli} et al.}{{Morbidelli} et al.}{2008}]{Morbidelli2008}
{Morbidelli} A., {Crida} A., {Masset} F., Nelson R. P., 2008, A\&A, 478, 929

\bibitem[\protect\citeauthoryear{{Monaghan}}{{Monaghan}}{1992}]{Monaghan1992}
{Monaghan} J. J., 1992, \araa, 30, 543

\bibitem[\protect\citeauthoryear{{Murray} \& {Holman}}{{Murray} \&
{Holman}}{1999}]{Murray1999}
{Murray} N., {Holman}, M., 1999, Science, 283, 1877

\bibitem[\protect\citeauthoryear{{Nagasawa} \& {Lin}}{{Nagasawa} \&
{Lin}}{2005}]{NagasawaLin2005} {Nagasawa} M., {Lin} D. N. C., 2005,
\apj, 532, 1140

\bibitem[\protect\citeauthoryear{{Nagasawa} et~al.}{{Nagasawa}
et~al.}{2005}]{Nagasawaetal2005} {Nagasawa} M., {Lin} D. N. C.,
{Thommes} E., 2005, \apj, 635, 578

\bibitem[\protect\citeauthoryear{{Novak} et~al.}{{Novak}
et~al.}{2003}]{Novak2003} {Novak} G. S., {Lai} D., {Lin} D. N. C.,
2003, in Scientific Frontiers in Research on Extrasolar Planets, ed.
{Deming} D. \& {Seager} S., (San Francisco: ASP), ASP conf. Ser., 294, p. 177


\bibitem[\protect\citeauthoryear{{Ogihara} \& {Ida}}{{Ogihara} \&
{Ida}}{2009}]{Ogihara09} {Ogihara} M., \& {Ida} S., 2009, \apj, 699,
824

\bibitem[\protect\citeauthoryear{{Ogilvie} \& {Lin}}{{Ogilvie} \&
{Lin}}{2004}]{Ogilvie2004} {Ogilvie} G. I., {Lin} D. N. C., 2004, \apj, 610,
477

\bibitem[\protect\citeauthoryear{{Paardekooper} \& {Papaloizou}}{{Paardekooper}
\& {Papaloizou}}{2008}]{Paardekooper2008} {Paardekooper} S. J.,
{Papaloizou} J. C. B., 2008, A\&A, 485, 877

\bibitem[\protect\citeauthoryear{{Paardekooper} et~al.}
{{Paardekooper}}{2010}]{Paarde2010}
{Paardekooper} S.-J., {Baruteau} C., {Crida} A., {Kley} W., 
2010, MNRAS, 401, 1950

\bibitem[\protect\citeauthoryear{{Papaloizou} \& {Terquem}}{{Papaloizou} \&
{Terquem}}{2006}]{Papaloizou2006} {Papaloizou} J. C. B., {Terquem} C., 2006,
Rep. Prog. Phys., 69, 119

\bibitem[\protect\citeauthoryear{{Pollack} et~al.}{{Pollack}
et~al.}{1996}]{Pollack1996} {Pollack} J. B., {Hubickyj} O., {Bodenheimer} P.,
{Lissauer} J. J., {Podolak} M., {Greenzweig} Y., 1996, Icarus, 124, 62

\bibitem[\protect\citeauthoryear{{Queloz} et~al.}{{Queloz}
et~al.}{2009}]{Queloz2009} {Queloz} D., Bouchy F., Moutou C., Hatzes, A.
et al. 2009, \aa, 506, 303

\bibitem[\protect\citeauthoryear{{Rafikov}}{{Rafikov}}{2006}]{Rafikov2006}
{Rafikov} R. R., 2006, \apj, 648, 666

\bibitem[\protect\citeauthoryear{{Safronov}}{{Safronov}}{1969}]{Safronov1969}
{Safronov} V. S., 1969, Evolution of the Protoplanetary Cloud and Formation of
the Earth and Planets (Moscow: Nauka Press)

\bibitem[\protect\citeauthoryear{{Sato} et al.}{{Sato} et al.}{2005}]{Sato2005}
{Sato} B., {Fischer} D. A., {Henry} G. W. et al., 2005, ApJ, 633, 465

\bibitem[\protect\citeauthoryear{{Saumon} et~al.}{{Saumon}
et~al.}{1995}]{Saumon1995} {Saumon} D., {Chabrier} G., {van Horn} H. M., 1995,
\apjs, 99, 713

\bibitem[\protect\citeauthoryear{{Schlaufman} et~al.}{{Schlaufman}
et~al.}{2009}]{Schlaufman2009} {Schlaufman} K., {Lin} D.N.C., {Ida}
S., 2009, \apj, 691, 1322

\bibitem[\protect\citeauthoryear{{Shiraishi} \& {Ida}}{{Shiraishi} \& {Ida}}{2008}]
{ShiraishiIda2008} {Shiraishi} M., {Ida} S., 2008, \apj, 684, 1416

\bibitem[\protect\citeauthoryear{{Showman} \& {Guillot}}{{Showman} \& {Guillot}}{2002}]
{Showman2002} {Showman} A. P., {Guillot} T., 2002, A\&A, 385, 166



\bibitem[Slattery et al.(1992)]{Slattery_etal_1992} Slattery, W.~L., Benz, 
W., \& Cameron, A.~G.~W.\ 1992, Icarus, 99, 167 



\bibitem[\protect\citeauthoryear{{Stevenson}}{{Stevenson}}{1976}]{Stevenson1976}
{Stevenson} D. J., 1976, Ph.D. thesis. Cornell University

\bibitem[\protect\citeauthoryear{{Stevenson}}{{Stevenson}}{1982}]{Stevenson1982}
{Stevenson} D. J., 1982, P\&SS, 30, 755

\bibitem[\protect\citeauthoryear{{Supulver} \& {Lin}}{{Supulver} \&{Lin}}{2000}]
{Supulver2000} {Supulver} K. D., {Lin} D. N. C., 2000, Icarus, 146, 525 

\bibitem[\protect\citeauthoryear{{Tanigawa} \& {Watanabe}}{{Tanigawa} \&
{Watanabe}}{2002}]{Tanigawa2002} {Tanigawa} T.,{Watanabe} S. I, 2002,
\apj, 580, 506

\bibitem[\protect\citeauthoryear{{Tanaka} et~al.}{{Tanaka}
et~al.}{1996}]{Tanaka1996} {Tanaka} H., {Inaba} S., {Nakazawa} K.,
1996, Icarus, 123, 450

\bibitem[\protect\citeauthoryear{{Terquem} \& {Papaloizou}}{{Terquem} \& {Papaloizou}}{2007}]{Terquem2007}
{Terquem} C., {Papaloizou} J.~C.~B., 2007, \apj, 654, 1110

\bibitem[\protect\citeauthoryear{{Thommes} et~al.}{{Thommes}
et~al.}{2008}]{Thommes2008} {Thommes} E., {Nagasawa} M., {Lin} D. N. C.,
2008, \apj, 676, 728

\bibitem[\protect\citeauthoryear{{Tillotson}}{{Tillotson}}{1962}]{Tillotson1962} {Tillotson} J.
H., 1962, General Atomic Reports, GA-3216

\bibitem[Tsiganis et al.(2005)]{Tsiganis2005} Tsiganis, K., Gomes, 
R., Morbidelli, A., \& Levison, H.~F.\ 2005, \nat, 435, 459 




\bibitem[\protect\citeauthoryear{{Ward}}{{Ward}}{1981}]{Ward1981} {Ward} W.
R., 1981, Icarus, 47, 234

\bibitem[\protect\citeauthoryear{{Ward}}{{Ward}}{1984}]{Ward1984}
{Ward} W. R., 1984, in Planetary rings, eds. Greenberg R. \& Brahic A. (Tucson: Univ. of
Arizona Press), 660

\bibitem[\protect\citeauthoryear{{Ward}}{{Ward}}{1997}]{Ward1997}
{Ward} W. R., 1997, Icarus, 126, 261

\bibitem[\protect\citeauthoryear{{Weidenschilling}}{{Weidenschilling}}{2004}]{Weidenschilling2004}
{Weidenschilling} S.~J., 2004, in Comets II, eds. Festou M.~C.,
Keller H.~U., \& {Weaver}, (Tucson: Univ. of Arizona Press), 97


\bibitem[Wetherill(1985)]{Wetherill_1985} Wetherill, G.~W.\ 1985, 
Science, 228, 877 


\bibitem[\protect\citeauthoryear{{Williams} \& {Wetherill}}{{Williams} \&
{Wetherill}}{1994}]{Williams1994} {Williams} D.~R, {Wetherill} G.~W., 1994,
Icarus, 107, 117

\bibitem[\protect\citeauthoryear{{Wuchterl}}{{Wuchterl}}{1991}]{Wuchterl1991}
{Wuchterl} G., 1991, Icarus, 91, 39

\bibitem[\protect\citeauthoryear{{Wuchterl} et~al.}{{Wuchterl}
et~al.}{2000}]{Wuchterl2000} {Wuchterl} G., {Guillot} T., {Lissauer} J.~J.,
2000, in Protostars and Planets IV, ed. {Mannings} V., {Boss} A. P.,
{Russell} S. S., (Tucson: Univ. Arizona Press), 1081

\bibitem[\protect\citeauthoryear{{Yu} \& {Tremaine}}{{Yu} \&
{Tremaine}}{2001}]{Yu2001} {Yu} Q., {Tremaine} S., 2001, \aj, 121, 1736

\bibitem[\protect\citeauthoryear{{Zhang} \& {Hamilton}}{{Zhang} \& {Hamilton}}{2007}]{Zhang2007}
{Zhang} K. {Hamilton} D. P., 2007, Icarus, 188, 386

\bibitem[\protect\citeauthoryear{{Zhou} et al.}{{Zhou} et al.}{2005}]{Zhou2005}
{Zhou} J.-L., {Aarseth} S. J., {Lin} D. N. C., {Nagasawa} M., 2005, ApJ, 631, L85

\bibitem[\protect\citeauthoryear{{Zhou} \& {Lin}}{{Zhou} \&
{Lin}}{2007}]{Zhou2007} {Zhou} J.-L., {Lin}, D. N. C., 2007, \apj, 666, 447

\bibitem[\protect\citeauthoryear{{Zhou} et al.}{{Zhou} et al.}{2007}]{Zhou2007b}
{Zhou} J.-L., {Lin} D. N. C., {Sun} Y.-S., 2007, ApJ, 666, 423

\bibitem[\protect\citeauthoryear{{Zhou} \& {Lin}}{{Zhou} \&
{Lin}}{2008}]{Zhou2008} {Zhou} J.-L., {Lin}, D. N. C., 2008, in IAU Symposium
249, Exoplanets: Detection, Formation and Dynamics, ed. Y.-S. Sun, S.
Ferraz-Mello, \& J.-L. Zhou, 285

\end{thebibliography}
\end{document}